\pdfoutput=1
\documentclass[12pt,a4paper]{article}

\usepackage{ifthen} 
\newboolean{pdflatex}
\setboolean{pdflatex}{true} 

\newboolean{articletitles}
\setboolean{articletitles}{true} 

\newboolean{uprightparticles}
\setboolean{uprightparticles}{false} 

\def\paperauthors{LHCb collaboration} 
\def\paperasciititle{Observation of open-charm tetraquark candidate Tcs0(2870)* in the B- -> D-D~0 KS decay} 
\def\papertitle{Observation of the open-charm tetraquark candidate $T_{cs 0}^{*}(2870)^0$ in the $B^- \rightarrow D^- D^0 K_\mathrm{S}^0$ decay} 
\def\paperkeywords{{High Energy Physics}, {LHCb}} 
\def\papercopyright{\the\year\ CERN for the benefit of the LHCb collaboration} 
\def\paperlicence{CC BY 4.0 licence}
\def\paperlicenceurl{https://creativecommons.org/licenses/by/4.0/}

\newif\ifEnableSectionTOCLinks
\EnableSectionTOCLinksfalse 

\usepackage[top=1in, bottom=1.25in, left=1in, right=1in]{geometry}

\usepackage{microtype}
\usepackage{lineno}  
\usepackage{xspace} 
\usepackage{caption} 

\usepackage{graphicx}  
\usepackage{color}
\usepackage{colortbl}
\graphicspath{{./figs/}} 

\usepackage{amsmath} 
\usepackage{amssymb}
\usepackage{amsfonts}
\usepackage{upgreek} 

\newcommand*\patchAmsMathEnvironmentForLineno[1]{%
\expandafter\let\csname old#1\expandafter\endcsname\csname #1\endcsname
\expandafter\let\csname oldend#1\expandafter\endcsname\csname
end#1\endcsname
 \renewenvironment{#1}%
   {\linenomath\csname old#1\endcsname}%
   {\csname oldend#1\endcsname\endlinenomath}%
}
\newcommand*\patchBothAmsMathEnvironmentsForLineno[1]{%
  \patchAmsMathEnvironmentForLineno{#1}%
  \patchAmsMathEnvironmentForLineno{#1*}%
}
\AtBeginDocument{%
\patchBothAmsMathEnvironmentsForLineno{equation}%
\patchBothAmsMathEnvironmentsForLineno{align}%
\patchBothAmsMathEnvironmentsForLineno{flalign}%
\patchBothAmsMathEnvironmentsForLineno{alignat}%
\patchBothAmsMathEnvironmentsForLineno{gather}%
\patchBothAmsMathEnvironmentsForLineno{multline}%
\patchBothAmsMathEnvironmentsForLineno{eqnarray}%
}

\usepackage{hyperxmp}

\usepackage[pdftex,
            pdfauthor={\paperauthors},
            pdftitle={\paperasciititle},
            pdfkeywords={\paperkeywords},
            pdfcopyright={Copyright (C) \papercopyright},
            pdflicenseurl={\paperlicenceurl}]{hyperref}

\usepackage[colorinlistoftodos,textsize=scriptsize]{todonotes}

\usepackage[bottom,flushmargin,hang,multiple]{footmisc}

\usepackage[all]{hypcap} 

\usepackage{xspace} 
\usepackage{upgreek}

\def\lhcb   {\mbox{LHCb}\xspace}

\def\MagUp {\mbox{\em Mag\kern -0.05em Up}\xspace}

\ifthenelse{\boolean{uprightparticles}}%
{

 \def\Ppi         {\ensuremath{\uppi}\xspace}

 \def\PDelta      {\ensuremath{\Delta}\xspace}                 
 \def\PXi         {\ensuremath{\Xi}\xspace}                 
 \def\PLambda     {\ensuremath{\Lambda}\xspace}                 
 \def\PSigma      {\ensuremath{\Sigma}\xspace}                 
 \def\POmega      {\ensuremath{\Omega}\xspace}                 
 \def\PUpsilon    {\ensuremath{\Upsilon}\xspace}
 \let\oldPi\Pi
 \def\PPi         {\ensuremath{\oldPi}\xspace}

 \def\PB      {\ensuremath{\mathrm{B}}\xspace}                 
                  
 \def\PD      {\ensuremath{\mathrm{D}}\xspace}

 \def\PK      {\ensuremath{\mathrm{K}}\xspace}

 \def\Pb      {\ensuremath{\mathrm{b}}\xspace}                 
 \def\Pc      {\ensuremath{\mathrm{c}}\xspace}                 
 \def\Pd      {\ensuremath{\mathrm{d}}\xspace}

 \def\Pi      {\ensuremath{\mathrm{i}}\xspace}

 \def\Ps      {\ensuremath{\mathrm{s}}\xspace}                 
                  
 \def\Pu      {\ensuremath{\mathrm{u}}\xspace}

 \def\thebaroffset{0.0em}
}
{

 \def\Ppi         {\ensuremath{\pi}\xspace}

 \mathchardef\PDelta="7101
 \mathchardef\PXi="7104
 \mathchardef\PLambda="7103
 \mathchardef\PSigma="7106
 \mathchardef\POmega="710A
 \mathchardef\PUpsilon="7107
 \mathchardef\PPi="7105
                  
 \def\PB      {\ensuremath{B}\xspace}                 
                  
 \def\PD      {\ensuremath{D}\xspace}

 \def\PK      {\ensuremath{K}\xspace}

 \def\Pb      {\ensuremath{b}\xspace}                 
 \def\Pc      {\ensuremath{c}\xspace}                 
 \def\Pd      {\ensuremath{d}\xspace}

 \def\Pi      {\ensuremath{i}\xspace}

 \def\Ps      {\ensuremath{s}\xspace}                 
                  
 \def\Pu      {\ensuremath{u}\xspace}

 \def\thebaroffset{0.18em}
}
\newcommand{\offsetoverline}[2][\thebaroffset]{\kern #1\overline{\kern -#1 #2}}%

\makeatletter
\ifcase \@ptsize \relax
  \newcommand{\miniscule}{\@setfontsize\miniscule{4}{5}}
\or
  \newcommand{\miniscule}{\@setfontsize\miniscule{5}{6}}
\or
  \newcommand{\miniscule}{\@setfontsize\miniscule{5}{6}}
\fi
\makeatother

\DeclareRobustCommand{\optbar}[1]{\shortstack{{\miniscule (\rule[.5ex]{1.25em}{.18mm})}
  \\ [-.7ex] $#1$}}

\def\uquark    {{\ensuremath{\Pu}}\xspace}
\def\uquarkbar {{\ensuremath{\overline \uquark}}\xspace}

\def\dquark    {{\ensuremath{\Pd}}\xspace}
\def\dquarkbar {{\ensuremath{\overline \dquark}}\xspace}

\def\squark    {{\ensuremath{\Ps}}\xspace}
\def\squarkbar {{\ensuremath{\overline \squark}}\xspace}

\def\cquark    {{\ensuremath{\Pc}}\xspace}
\def\cquarkbar {{\ensuremath{\overline \cquark}}\xspace}

\def\bquark    {{\ensuremath{\Pb}}\xspace}

\def\pion   {{\ensuremath{\Ppi}}\xspace}

\def\pip    {{\ensuremath{\pion^+}}\xspace}
\def\pim    {{\ensuremath{\pion^-}}\xspace}

\def\pimp   {{\ensuremath{\pion^\mp}}\xspace}

\def\kaon    {{\ensuremath{\PK}}\xspace}
\def\Kbar    {{\ensuremath{\offsetoverline{\PK}}}\xspace}

\def\KorKbar {\kern \thebaroffset\optbar{\kern -\thebaroffset \PK}{}\xspace}

\def\Kzb     {{\ensuremath{\Kbar{}^0}}\xspace}
\def\Kp      {{\ensuremath{\kaon^+}}\xspace}
\def\Km      {{\ensuremath{\kaon^-}}\xspace}

\def\KS      {{\ensuremath{\kaon^0_{\mathrm{S}}}}\xspace}

\def\Kstarb  {{\ensuremath{\Kbar{}^*}}\xspace}

\def\Dbar    {{\ensuremath{\offsetoverline{\PD}}}\xspace}
\def\D       {{\ensuremath{\PD}}\xspace}

\def\DorDbar {\kern \thebaroffset\optbar{\kern -\thebaroffset \PD}\xspace}
\def\Dz      {{\ensuremath{\D^0}}\xspace}
\def\Dzb     {{\ensuremath{\Dbar{}^0}}\xspace}
\def\Dp      {{\ensuremath{\D^+}}\xspace}
\def\Dm      {{\ensuremath{\D^-}}\xspace}

\def\DpDm    {\ensuremath{\Dp {\kern -0.16em \Dm}}\xspace}
\def\Dstar   {{\ensuremath{\D^*}}\xspace}

\def\Dstarm  {{\ensuremath{\D^{*-}}}\xspace}

\def\Dsp     {{\ensuremath{\D^+_\squark}}\xspace}
\def\Dsm     {{\ensuremath{\D^-_\squark}}\xspace}

\def\B       {{\ensuremath{\PB}}\xspace}

\def\BorBbar {\kern \thebaroffset\optbar{\kern -\thebaroffset \PB}\xspace}
\def\Bz      {{\ensuremath{\B^0}}\xspace}

\def\Bd      {{\ensuremath{\B^0}}\xspace}

\def\BdorBdbar {\kern \thebaroffset\optbar{\kern -\thebaroffset \Bd}\xspace}
\def\Bu      {{\ensuremath{\B^+}}\xspace}
\def\Bub     {{\ensuremath{\B^-}}\xspace}
\def\Bp      {{\ensuremath{\Bu}}\xspace}
\def\Bm      {{\ensuremath{\Bub}}\xspace}

\def\Bs      {{\ensuremath{\B^0_\squark}}\xspace}

\def\BsorBsbar {\kern \thebaroffset\optbar{\kern -\thebaroffset \Bs}\xspace}

\def\Y#1S{\ensuremath{\PUpsilon{(#1S)}}\xspace}

\def\LorLbar     {\kern \thebaroffset\optbar{\kern -\thebaroffset \PLambda}\xspace}

\def\BF         {{\ensuremath{\mathcal{B}}}\xspace}
\def\BR         {\BF}

\newcommand{\decay}[2]{\ensuremath{#1\!\to #2}\xspace} 

\def\to                 {\ensuremath{\rightarrow}\xspace}

\def\AT#1     {\ensuremath{A_{\mathrm{T}}^{#1}}\xspace}           

\def\C#1      {\ensuremath{\mathcal{C}_{#1}}\xspace}                       
\def\Cp#1     {\ensuremath{\mathcal{C}_{#1}^{'}}\xspace}                    
\def\Ceff#1   {\ensuremath{\mathcal{C}_{#1}^{\mathrm{(eff)}}}\xspace}        
\def\Cpeff#1  {\ensuremath{\mathcal{C}_{#1}^{'\mathrm{(eff)}}}\xspace}       
\def\Ope#1    {\ensuremath{\mathcal{O}_{#1}}\xspace}                       
\def\Opep#1   {\ensuremath{\mathcal{O}_{#1}^{'}}\xspace}                    

\newcommand{\aunit}[1]{\ensuremath{\text{\,#1}}}       

\newcommand{\tev}{\aunit{Te\kern -0.1em V}\xspace}
\newcommand{\gev}{\aunit{Ge\kern -0.1em V}\xspace}
\newcommand{\mev}{\aunit{Me\kern -0.1em V}\xspace}
\newcommand{\kev}{\aunit{ke\kern -0.1em V}\xspace}
\newcommand{\ev}{\aunit{e\kern -0.1em V}\xspace}
 
\newcommand{\mevc}{\ensuremath{\aunit{Me\kern -0.1em V\!/}c}\xspace}
\newcommand{\gevc}{\ensuremath{\aunit{Ge\kern -0.1em V\!/}c}\xspace}
\newcommand{\mevcc}{\ensuremath{\aunit{Me\kern -0.1em V\!/}c^2}\xspace}
\newcommand{\gevcc}{\ensuremath{\aunit{Ge\kern -0.1em V\!/}c^2}\xspace}

\def\fb   {\ensuremath{\aunit{fb}}\xspace}
\def\invfb   {\ensuremath{\fb^{-1}}\xspace}

\def\gsim{{~\raise.15em\hbox{$>$}\kern-.85em
          \lower.35em\hbox{$\sim$}~}\xspace}
\def\lsim{{~\raise.15em\hbox{$<$}\kern-.85em
          \lower.35em\hbox{$\sim$}~}\xspace}

\newcommand{\Real}{\ensuremath{\mathcal{R}e}\xspace}
\newcommand{\Imag}{\ensuremath{\mathcal{I}m}\xspace}

\def\pt         {\ensuremath{p_{\mathrm{T}}}\xspace}

\def\tell1  {TELL1\xspace}
\def\ukl1   {UKL1\xspace}

\newcommand{\eg}{\mbox{\itshape e.g.}\xspace}
\newcommand{\ie}{\mbox{\itshape i.e.}\xspace}

\newcommand{\etc}{\mbox{\itshape etc.}\xspace}

\newcommand{\lhcborcid}[1]{\href{https://orcid.org/#1}{\hspace*{0.1em}\raisebox{-0.45ex}{\includegraphics[width=1em]{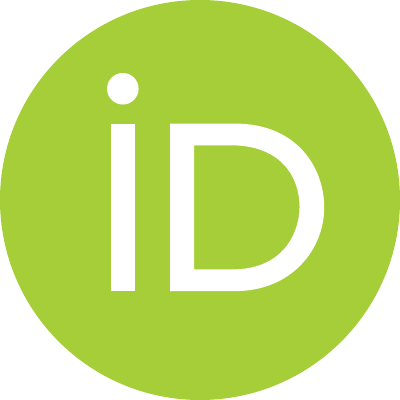}}}}

\def\myTcsZero  {{\ensuremath{T_{\cquark\squark 0}^{*}(2870)^0}}\xspace}
\def\myTcsOne  {{\ensuremath{T_{\cquark\squark 1}^{*}(2900)^0}}\xspace}
\def\myTcsZeroShort  {{\ensuremath{T_{\cquark\squark 0}^{*0}}}\xspace}
\def\myTcsOneShort  {{\ensuremath{T_{\cquark\squark 1}^{*0}}}\xspace}
\def\myTcsZeroOne  {{\ensuremath{T_{\cquark\squark 0,1}^{*0}}}\xspace}
\def\TcsGeneric  {{\ensuremath{T_{\cquark\squark}^{*0}}}\xspace}
\def\myBmDecay {{\ensuremath{\decay{\Bm}{\Dm\Dz\KS}}}\xspace}

\def\Dsstar  {{\ensuremath{\D_{sJ}^{*-}}}\xspace}

\hypersetup{
  colorlinks   = true, 
  urlcolor     = blue, 
  linkcolor    = blue, 
  citecolor    = red   
}

\ifEnableSectionTOCLinks
    \usepackage[explicit]{titlesec} 
    
    \let\oldcontentsline\contentsline
    \renewcommand

    \titleformat{\section}{\normalfont\Large\bf}{\hyperlink{tocsection.\thesection}{{\thesection} \parbox[t]{\dimexpr\textwidth-1pc}{#1}}}{1pc}{}

    \titleformat{\subsection}{\normalfont\bf}{\hyperlink{tocsubsection.\thesubsection}{{\thesubsection} \parbox[t]{\dimexpr\textwidth-1pc}{#1}}}{1pc}{}

    \titleformat{name=\section,numberless}[display]{}{}{0pt}{\normalfont\Huge\bfseries #1}
\fi

\usepackage{cite} 
\usepackage{mciteplus}

\usepackage{longtable} 

\usepackage{comment}
\usepackage{tikz-feynman}
\usepackage{feynmp-auto}
\usepackage{multicol}
\usepackage{multirow}

\begin{document}

\renewcommand{\thefootnote}{\fnsymbol{footnote}}
\setcounter{footnote}{1}


\begin{titlepage}
\pagenumbering{roman}

\vspace*{-1.5cm}
\centerline{\large EUROPEAN ORGANIZATION FOR NUCLEAR RESEARCH (CERN)}
\vspace*{1.5cm}
\noindent
\begin{tabular*}{\linewidth}{lc@{\extracolsep{\fill}}r@{\extracolsep{0pt}}}
\ifthenelse{\boolean{pdflatex}}
{\vspace*{-1.5cm}\mbox{\!\!\!\includegraphics[width=.14\textwidth]{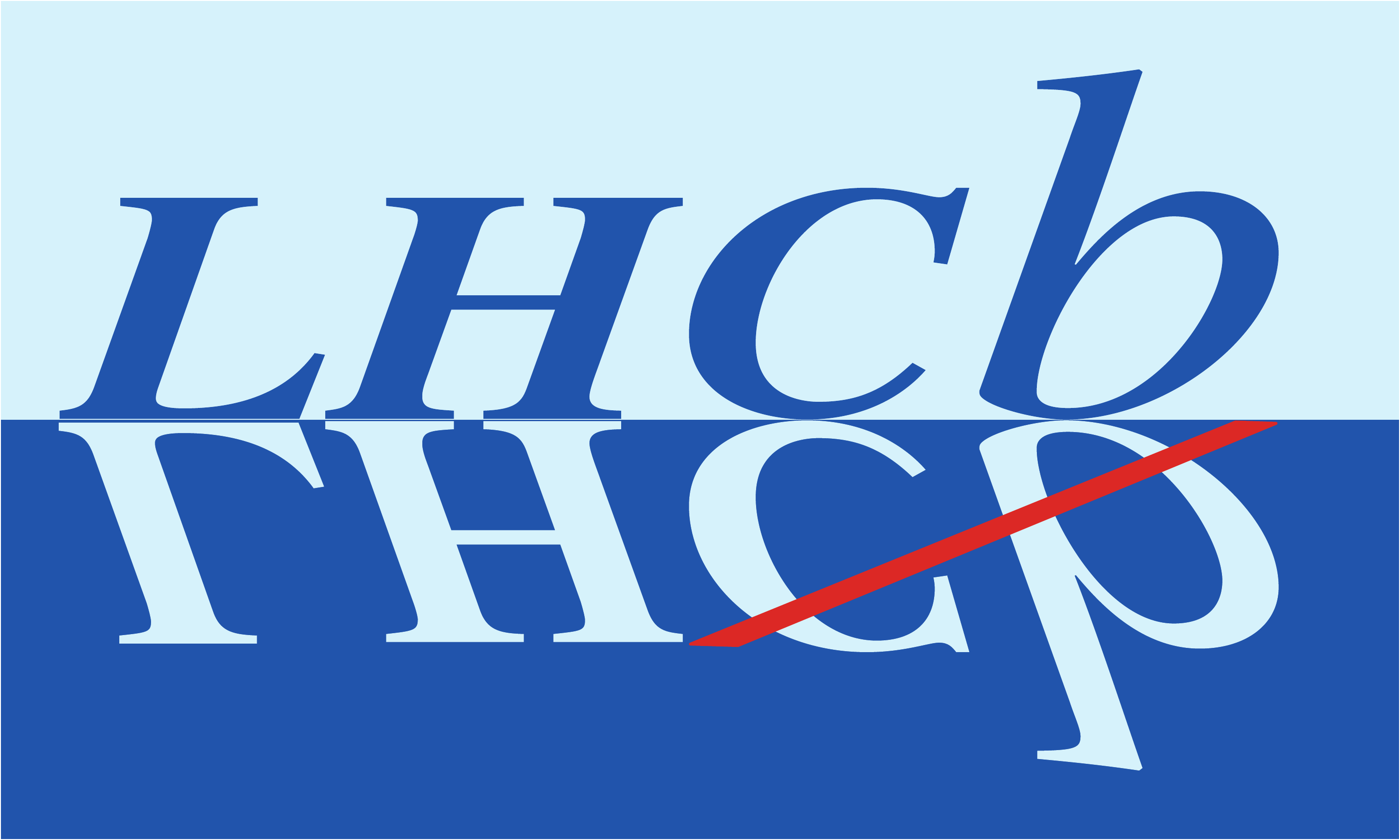}} & &}%
{\vspace*{-1.2cm}\mbox{\!\!\!\includegraphics[width=.12\textwidth]{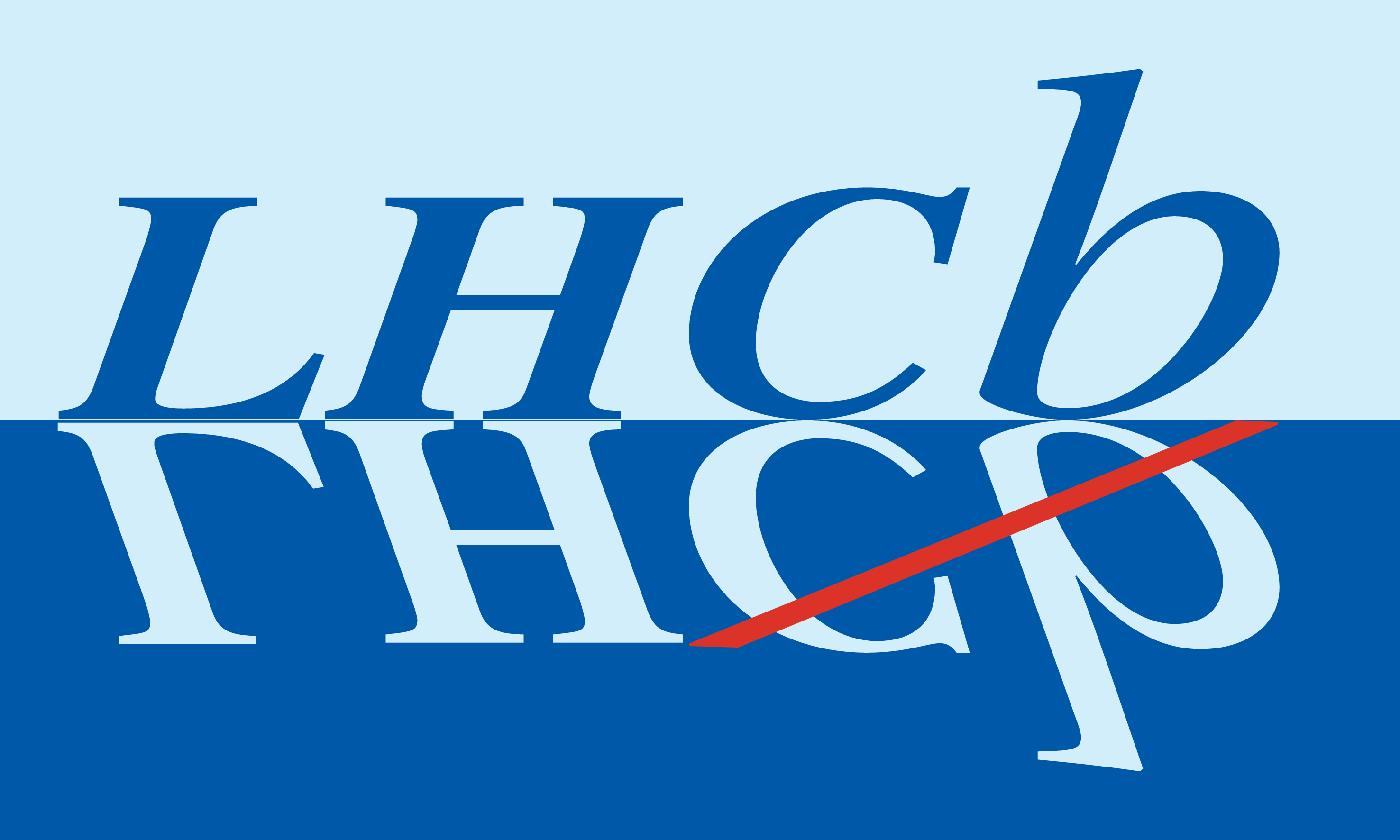}} & &}%
\\
 & & CERN-EP-2024-287 \\  
 & & LHCb-PAPER-2024-040 \\  
 & & March 12, 2025 \\ 
 & & \\
\end{tabular*}

\vspace*{1.0cm}

{\normalfont\bfseries\boldmath\huge
\begin{center}
  \papertitle 
\end{center}
}

\vspace*{2.0cm}

\begin{center}
\paperauthors\footnote{Authors are listed at the end of this paper.}
\end{center}

\vspace{\fill}

\begin{abstract}
  \noindent
An amplitude analysis of $B^-\rightarrow D^- D^0 K_\mathrm{S}^0$ decays is performed using proton-proton collision data, corresponding to an integrated luminosity of $9\mbox{\,fb}^{-1}$, collected with the \mbox{LHCb} detector at center-of-mass energies of 7, 8, and 13$\mathrm{\,Te\kern -0.1em V}$. 
A resonant structure of spin-parity $0^+$ is observed in the $D^0 K_\mathrm{S}^0$ invariant-mass spectrum with a significance of $5.3\,\sigma$. 
The mass and width of the state, modeled with a \mbox{Breit--Wigner} line shape, are determined to be $2883\pm11\pm8\mathrm{\,Me\kern -0.1em V\!/}c^2$ and $87_{-47}^{+22}\pm17\mathrm{\,Me\kern -0.1em V}$ respectively, where the first uncertainties are statistical and the second systematic. These properties and the quark content are consistent with those of the open-charm tetraquark candidate $T_{cs 0}^{*}(2870)^0$ observed previously in the $D^+ K^-$ final state of the $B^-\rightarrow D^- D^+ K^-$ decay. 
This result  confirms the existence of the $T_{cs 0}^{*}(2870)^0$ state in a new decay mode.
The $T_{cs1}^{*}(2900)^0$ state, reported in the $B^-\rightarrow D^- D^+ K^-$ decay, is also searched for in the $D^0 K_\mathrm{S}^0$ invariant-mass spectrum of the $B^- \rightarrow D^- D^0 K_\mathrm{S}^0$ decay, without finding evidence for it.
\end{abstract}

\vspace*{2.0cm}

\begin{center}
  Published in \href{https://doi.org/10.1103/PhysRevLett.134.101901}{Phys. Rev. Lett. 134 (2025) 101901}
\end{center}

\vspace{\fill}

{\footnotesize 
\centerline{\copyright~\papercopyright. \href{\paperlicenceurl}{\paperlicence}.}}
\vspace*{2mm}

\end{titlepage}


\newpage
\setcounter{page}{2}
\mbox{~}
%
%
%
%


\renewcommand{\thefootnote}{\arabic{footnote}}
\setcounter{footnote}{0}

\cleardoublepage


\pagestyle{plain} 
\setcounter{page}{1}
\pagenumbering{arabic}


Hadrons composed of more than three quarks, referred to in this Letter as exotic states, play a unique role in understanding the confinement mechanism of the strong interaction~\cite{Liu:2019zoy,Guo:2017jvc}. 
In two decades of advances after the $\chi_{\cquark 1}(3872)$ discovery in 2003~\cite{Belle:2003nnu}, a variety of exotic candidates has been experimentally observed. 
These include the charmoniumlike tetraquark $T_{\cquark\cquarkbar 1}(3900)^+$~\cite{BESIII:2013ris, Belle:2013yex}, tetraquarks with two  or four charm quarks $T_{\cquark\cquark}(3875)^+$~\cite{LHCb-PAPER-2021-031, LHCb-PAPER-2021-032} and $T_{\cquark\cquark\cquarkbar\cquarkbar}(6900)^0$~\cite{LHCb-PAPER-2020-011, ATLAS:2023bft, CMS:2023owd}, respectively, and  pentaquark $P_{\cquark\cquarkbar}^+$ and  $P_{\cquark\cquarkbar\squark}^0$ candidates~\cite{LHCb-PAPER-2015-029, LHCb-PAPER-2019-014, LHCb-PAPER-2021-018, LHCb-PAPER-2020-039, LHCb-PAPER-2022-031}.  
In 2020, two states $\myTcsZero$ and $\myTcsOne$, with  spin-parity $J^P=0^+$ and $1^-$ respectively, and minimum quark content $\cquark\squark\uquarkbar\dquarkbar$, were observed by the \lhcb Collaboration in the $\Dp\Km$ final state of the $\decay{\Bm}{\Dm\Dp\Km}$ decay~\cite{LHCb-PAPER-2020-025,LHCb-PAPER-2020-024}. This was the first observation of manifestly exotic candidates with a single charm quark, opening a new avenue for understanding the quark-binding mechanism. 
The existence of these states was confirmed in the $\Dp\Km$ invariant-mass spectrum of the $\decay{\Bm}{\Dstarm\Dp\Km}$ decay~\cite{LHCb-PAPER-2023-047}. 
Charged open-charm tetraquarks were searched for in $\Bz\to\Dp\Dm\KS$ decays, but without any significant signal~\cite{LHCb-PAPER-2023-042}.
Recently, two new open-charm tetraquark candidates $T_{\cquark\squarkbar0}^*(2900)^{0/++}$, with $J^P=0^+$ and minimum quark content $\cquark\squarkbar\uquarkbar\dquark/\cquark\squarkbar\uquark\dquarkbar$, were observed in the $\Dsp\pimp$ final state of the $\decay{\Bz}{\Dzb\Dsp\pim}$ and $\decay{\Bp}{\Dm\Dsp\pip}$ decays~\cite{LHCb-PAPER-2022-026,LHCb-PAPER-2022-027}. Charge conjugation of states and decays is implied throughout this Letter.

Studies have been carried out to understand the nature of the  $\myTcsZero$ and $\myTcsOne$ structures, collectively referred to as $\TcsGeneric$ hereafter. 
They can be interpreted as compact tetraquarks~\cite{Karliner:2020vsi, He:2020jna, Guo:2021mja, Ozdem:2022ydv, Wang:2020xyc, Agaev:2021knl, Yang:2021izl, Wang:2020prk}  or $\PD^{(*)}\Kstarb$ hadronic molecules~\cite{Molina:2010tx, Molina:2020hde, Hu:2020mxp, Liu:2020nil, Kong:2021ohg, Wang:2021lwy, Xiao:2020ltm, Chen:2020aos, Chen:2021xlu, Huang:2020ptc}.
In either case, assuming isospin symmetry is respected, the $\TcsGeneric$ hadrons should decay to both $\Dp\Km$ and $\Dz\Kzb$ final states with similar rates~\cite{Huang:2020ptc,Burns:2020xne,Qin:2022nof}, as the two systems have the same minimum quark content. 
The two $\TcsGeneric$ structures can also be explained as kinematically generated singularities, originating from intermediate three-hadron loop diagrams~\cite{Burns:2020xne}.
In this scenario, the $\TcsGeneric$ structures produced in the $\Bm\to\Dm \TcsGeneric$ decay exhibit larger decay rates into the $\Dp\Km$ final state than into the $\Dz\Kzb$ final state~\cite{Burns:2020xne}.
Given the present limited experimental inputs, it is still difficult to establish which interpretation is more likely~\cite{Chen:2022asf}.

This Letter presents an amplitude analysis of the $\decay{\Bm}{\Dm\Dz\Kzb}$ decay, where the $\TcsGeneric$ resonances could be observed in the $\Dz\Kzb$ final state, and the $\Kzb$ meson is reconstructed in the $\KS$ mass eigenstate.
The analysis is performed using proton-proton ($pp$) collision data collected by the~\lhcb experiment, corresponding to an integrated luminosity of $9\invfb$ at center-of-mass energies of 7, 8 and 13\tev. 
The \lhcb detector~\cite{LHCb-DP-2008-001,LHCb-DP-2014-002} is a
single-arm forward spectrometer covering the pseudorapidity range $2 < \eta < 5$. 
The detector elements that are particularly relevant to this analysis include a silicon-strip vertex detector surrounding the $pp$ interaction region that allows \cquark and \bquark hadrons to be identified from their characteristically long flight distance, a high-precision tracking system to measure the momentum, $p$, of charged
particles, and two ring-imaging Cherenkov detectors to perform particle identification (PID) for charged hadrons.
The online event selection of $\myBmDecay$ decays is performed using a trigger, which consists of a hardware stage based on information from the calorimeter and muon
systems, followed by a software stage, which applies a full event
reconstruction.

In the off-line analysis, $\Bm$ candidates are selected by combining $\Dm$, $\Dz$ and $\KS$ candidates.
The $\Dm$ and $\KS$ mesons  are reconstructed in $\decay{\Dm}{\Kp\pim\pim}$  and $\decay{\KS}{\pip\pim}$ decays, respectively, while the $\Dz$ meson is formed according to either the $\decay{\Dz}{\Km\pip}$ or the $\decay{\Dz}{\Km\pip\pim\pip}$ decay.  
The final-state tracks are required to be positively identified as pions or kaons by the PID detectors, to have a large transverse momentum ($\pt$), and to be well separated from any $pp$ collision vertex (PV).
The $\Dm$, $\Dz$ and $\KS$ candidates should have reconstructed decay vertices of good fit quality and invariant masses consistent with known values from Ref.~\cite{PDG2024}. 
The reconstructed decay vertex of the $\Bm$ meson is required to have a good fit quality and be significantly displaced from  its associated PV, defined as the PV
 that aligns best with the flight direction of the $\Bm$ candidate.
To improve the experimental resolution on the reconstructed $\Bm$ invariant mass, $m_{\Dm\Dz\KS}$, a kinematic fit to the whole decay chain is performed,  with the $\Dm$, $\Dz$, and $\KS$ invariant masses constrained to their known values~\cite{PDG2024} and the $\Bm$ momentum constrained to point back to its associated PV~\cite{Hulsbergen:2005pu}. 
The background, dominated by random combinations of $\Dm$, $\Dz$, and $\KS$ candidates,
is further suppressed by a boosted decision tree~(BDT)~\cite{Breiman,AdaBoost}
implemented in the TMVA toolkit~\cite{Hocker:2007ht,*TMVA4}. 
The BDT classifier is trained using a sample of simulated $\mbox{\decay{\Bm}{\Dm\Dz\KS}}$ candidates as signal sample and candidates with \Bm mass exceeding $5500 \mevcc$ from data as background sample. Training variables include those characterizing the decay topology, particle transverse momenta, vertex fit quality and particle identification information for pions and kaons. 
The requirement on the BDT classifier maximizes the figure of merit \mbox{$N_S^2/(N_S+N_B)^{3/2}$}~\cite{LHCb-PAPER-2020-025}, where $N_S$ and $N_B$ represent the signal and background yields in the signal region, which is defined as $\pm20\mevcc$ around the known $\Bm$ mass~\cite{PDG2024}.
Possible physics backgrounds are investigated, notably the $\decay{B}{\Dstar \Dm\KS}$ decay with $\decay{\Dstar}{\Dz\pi/\gamma}$, where the $\pi/\gamma$ particle is not reconstructed, and the 
 $\decay{\Bm}{\Dz\KS\Kp\pim\pim}$ five-body decay. 
The former is shifted and well separated from the $\Bm$ signal, and the latter is suppressed to a negligible level by requiring that the vertices of the $\Dm$ and $\Bm$ candidates are well separated. 

A simulated sample of $\myBmDecay$ decays, generated with the software packages
described in Refs.~\cite{Sjostrand:2007gs, Lange:2001uf, Allison:2006ve, Muller:2018vny}, is used to model the effects of the detector acceptance and the imposed selection requirements. The \Bm transverse momentum and rapidity distribution, as well as the PID and trigger responses, are weighted to match the data.
The simulated decays are subjected to the same reconstruction and selection procedures as the data. 

\begin{figure}[!tbhp]
    \centering
    \includegraphics[width=0.49\columnwidth]{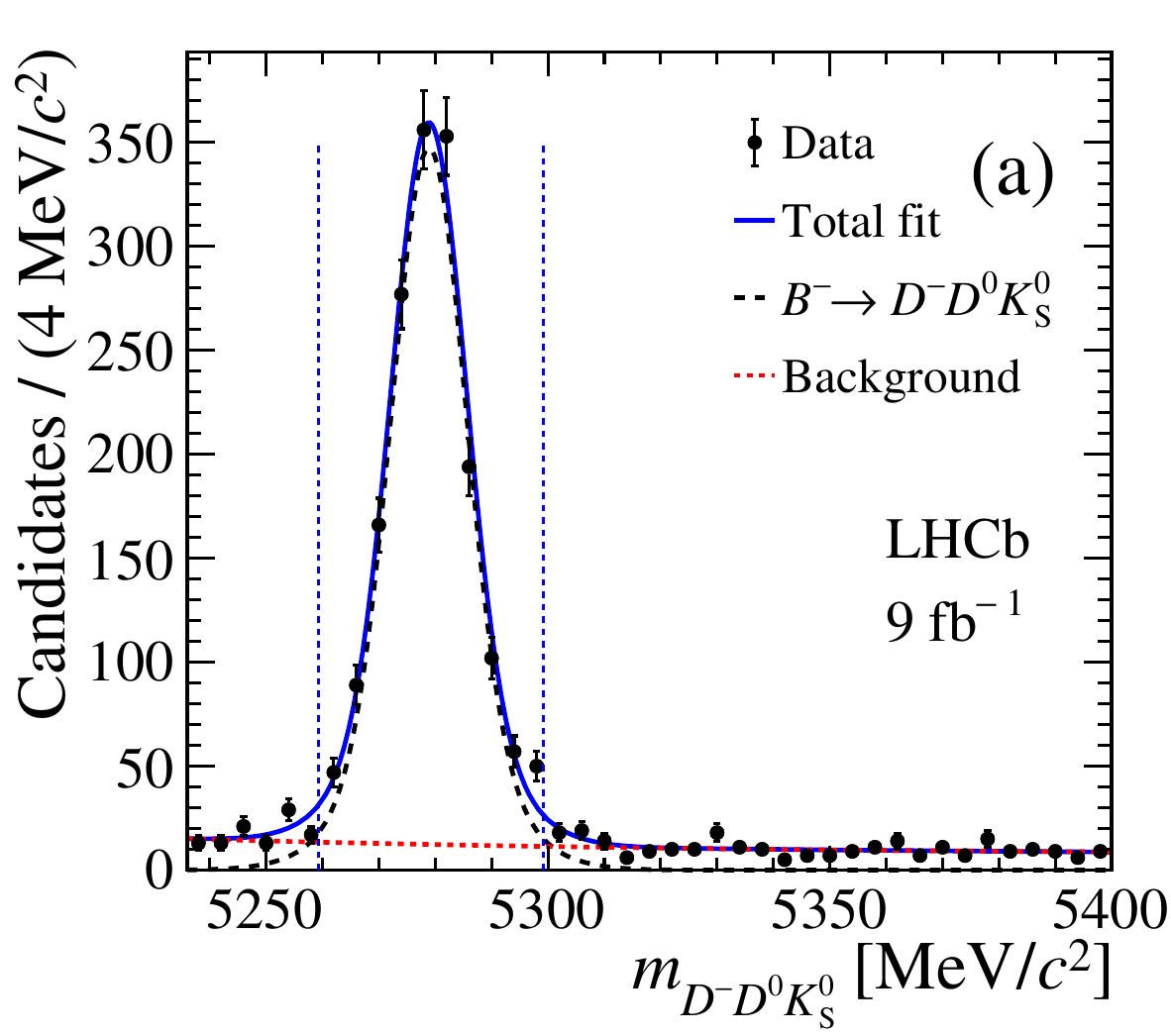}
    \includegraphics[width=0.49\columnwidth]{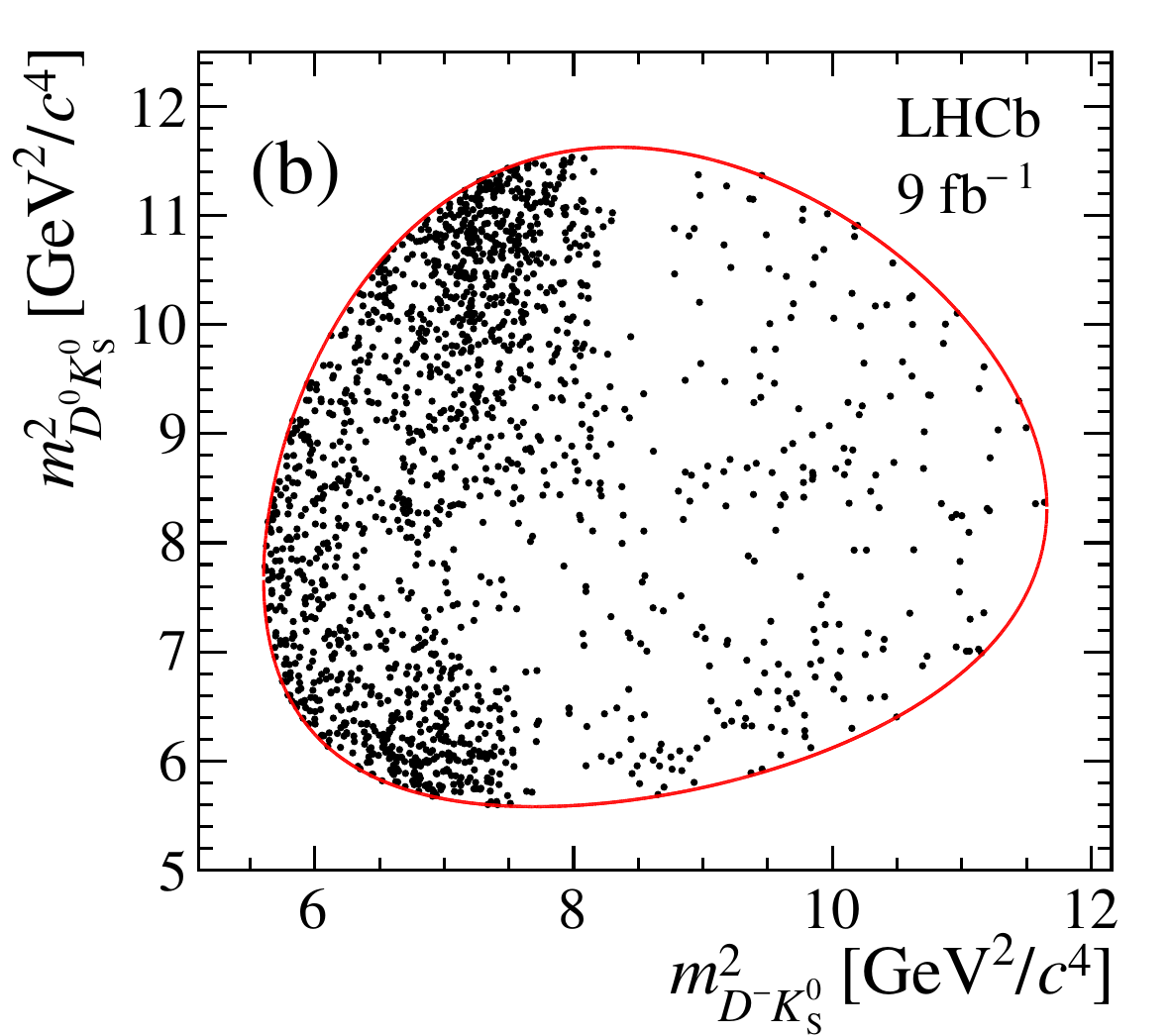}
    \caption{
    (a) Invariant-mass distribution of the $B^-$ candidates with the fit results also shown. The blue dashed lines illustrate the $B^-$ signal region corresponding to a signal yield of $1540\pm40$ with a purity of $(92.6\pm0.6)\%$.
    (b) Dalitz plot of $B^-$ candidates within the signal region. The red solid curve represents the kinematic boundary.}
    \label{fig:massfit}
\end{figure}

In order to extract the yield of the signal decay mode, an unbinned extended maximum-likelihood fit is performed to the $m_{\Dm\Dz\KS}$ distribution. 
The signal is described by a combination of two Gaussian functions with a shared mean value. 
The combinatorial background is modeled by an exponential function. 
The relative proportions of the two Gaussian functions and the ratio of their widths are fixed to values obtained in the simulated samples, while the remaining parameters are floated in the fit.
The $m_{\Dm\Dz\KS}$ distribution is shown in Fig.~\ref{fig:massfit}(a), together with the fit results. 
In the signal region, the yield and fraction of the $\myBmDecay$ signal decay are determined to be 
$1540\pm 40$ and $f_\text{s}=(92.6\pm 0.6)\%$, respectively. 

The Dalitz plot of the $\myBmDecay$ decay, represented by the squared two-body invariant masses $m^2_{\Dm\KS}$ and $m^2_{\Dz\KS}$, is shown in Fig.~\ref{fig:massfit}(b) for candidates in the $\Bm$ signal region. 
The Dalitz-plot variables are calculated with the masses of $\Bm$, $\Dm$, $\Dz$, and $\KS$ mesons constrained to their known values~\cite{PDG2024} and the $\Bm$ momentum constrained to point back to its associated PV~\cite{Hulsbergen:2005pu}.
Excited $\Dsm$ mesons, $\Dsstar$,  are clearly visible in the low $m^2_{\Dm\KS}$ region. 
An amplitude analysis is performed to investigate all resonance structures in the $\myBmDecay$ decays. 
The total probability density function (PDF) is described as the sum of those for the \Bm signal $\mathcal{P}_\text{s}$  and the combinatorial background $\mathcal{P}_\text{b}$ with the signal fraction $f_\text{s}$ fixed to the value measured by the fit to the $m_{\Dm\Dz\KS}$ mass spectrum. The PDF of the combinatorial background is determined using data in the mass region $5350<m_{\Dm\Dz\KS}<5565 \mevcc$. 
For the signal PDF, contributions from two interfering decay sequences are considered: \mbox{$\decay{\Bm}{\Dsstar(\to\Dm\KS) \Dz}$} and \mbox{$\decay{\Bm}{\TcsGeneric(\to\Dz\KS)\Dm }$}, collectively denoted as \mbox{$\decay{\Bm}{R(\to a b) c}$}.
For each decay sequence, various $\Dsstar$ or $\TcsGeneric$ components are investigated, and each component contributes  to the decay with an independent amplitude, $\mathcal{M}_R(m_{ab},\theta_{ab}|\vec{\omega})$, where $\theta_{ab}$ is the angle between the momenta of $a$ and the $\Bm$ in the $R$ rest frame, and the $m_{ab}$ variable is the invariant mass of the $ab$ system~\cite{LHCb-PAPER-2022-027}. 
The set of free parameters $\vec{\omega}$ includes the complex coupling for each amplitude and the masses and widths of unknown states.  
The distribution of candidates in the Dalitz plane is proportional to the squared total amplitude defined as the coherent sum of each amplitude  determined by the
helicity formalism~\cite{JACOB1959404}.
Accounting for the nonuniform experimental efficiency across the Dalitz plot, $\epsilon(m_{ab},\theta_{ab})$, the signal PDF is determined to be
\begin{equation*}
        \mathcal{P}_\text{s}(m_{ab}, \theta_{ab}|\vec{\omega})=
      \frac{\epsilon(m_{ab},\theta_{ab})}{I(\vec{\omega})}\left | \sum_R \mathcal{M}_R(m_{ab},\theta_{ab}|\vec{\omega}) \right | ^2,
\end{equation*}
where  $I(\vec{\omega})$ is a normalization factor, and the index $R$  runs over the considered contributions.
The efficiency is determined  in bins of the Dalitz plot using simulated $\myBmDecay$ decays.
The log-likelihood is
\begin{equation*}
 \ln\mathcal{L}=\sum_{j}\ln\left[ f_\text{s}\mathcal{P}_\text{s}(m_{ab}^j,\theta_{ab}^j|\vec{\omega}) +(1-f_\text{s})\mathcal{P}_\text{b}(m_{ab}^j,\theta_{ab}^j) \right],
\end{equation*}
where the $j$ index runs over the number of the $\Bm$ candidates in the signal region. 
Maximization of $\ln \mathcal{L}$  returns the estimated values of the unknown parameters.

\begin{figure*}[!t]
    \centering
    \includegraphics[width=0.32\columnwidth]{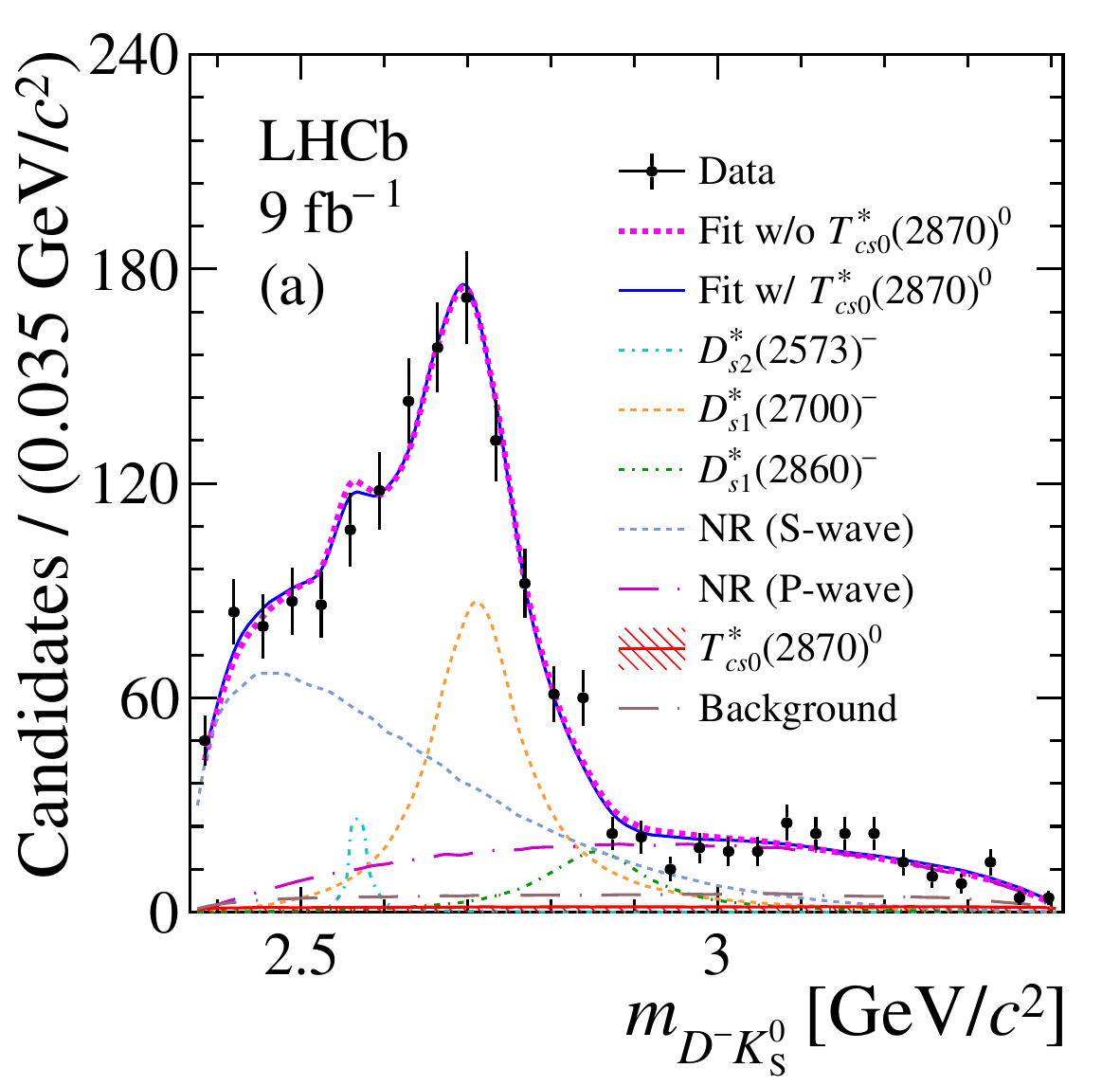}
    \includegraphics[width=0.32\columnwidth]{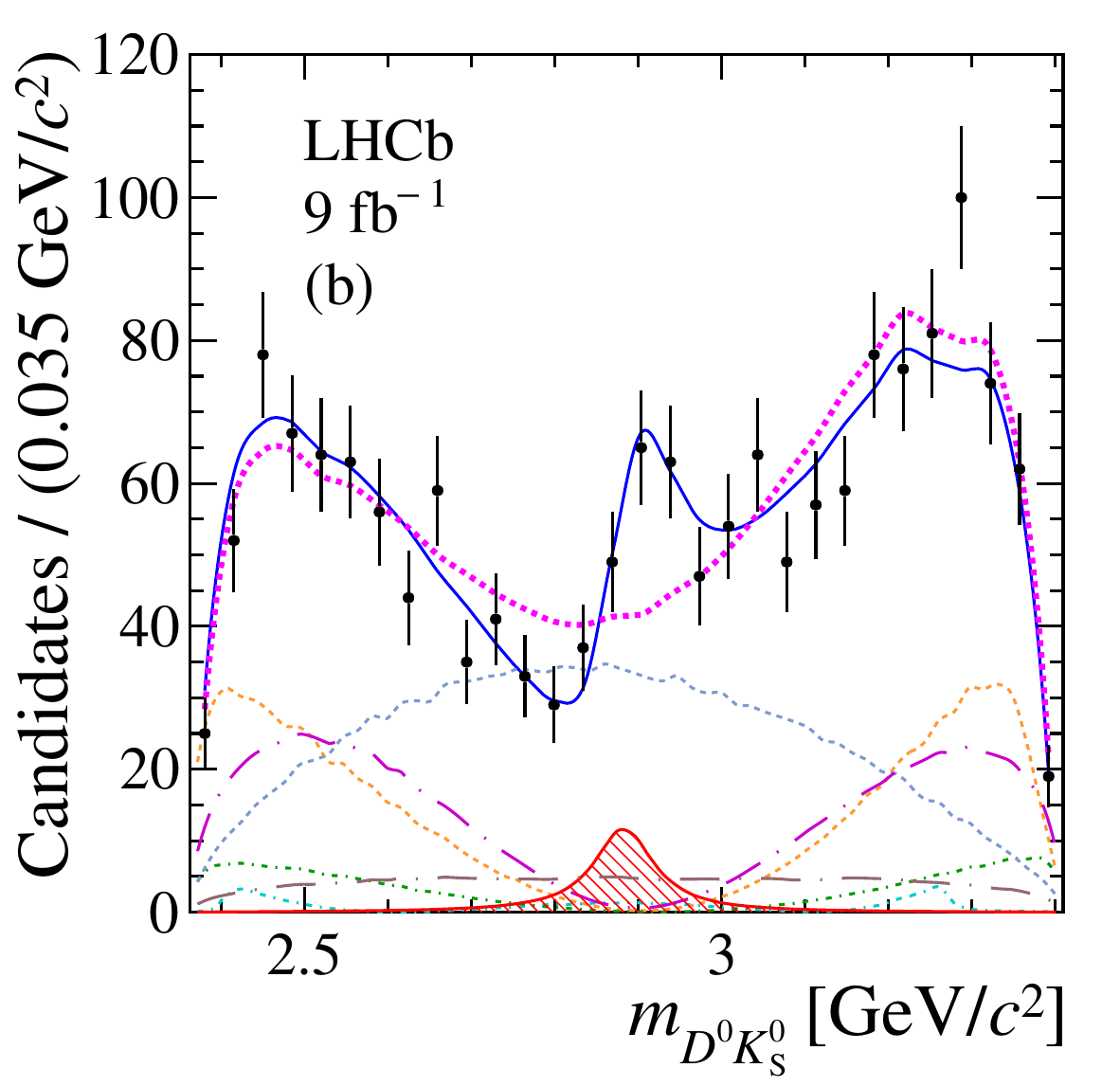}
    \includegraphics[width=0.32\columnwidth]{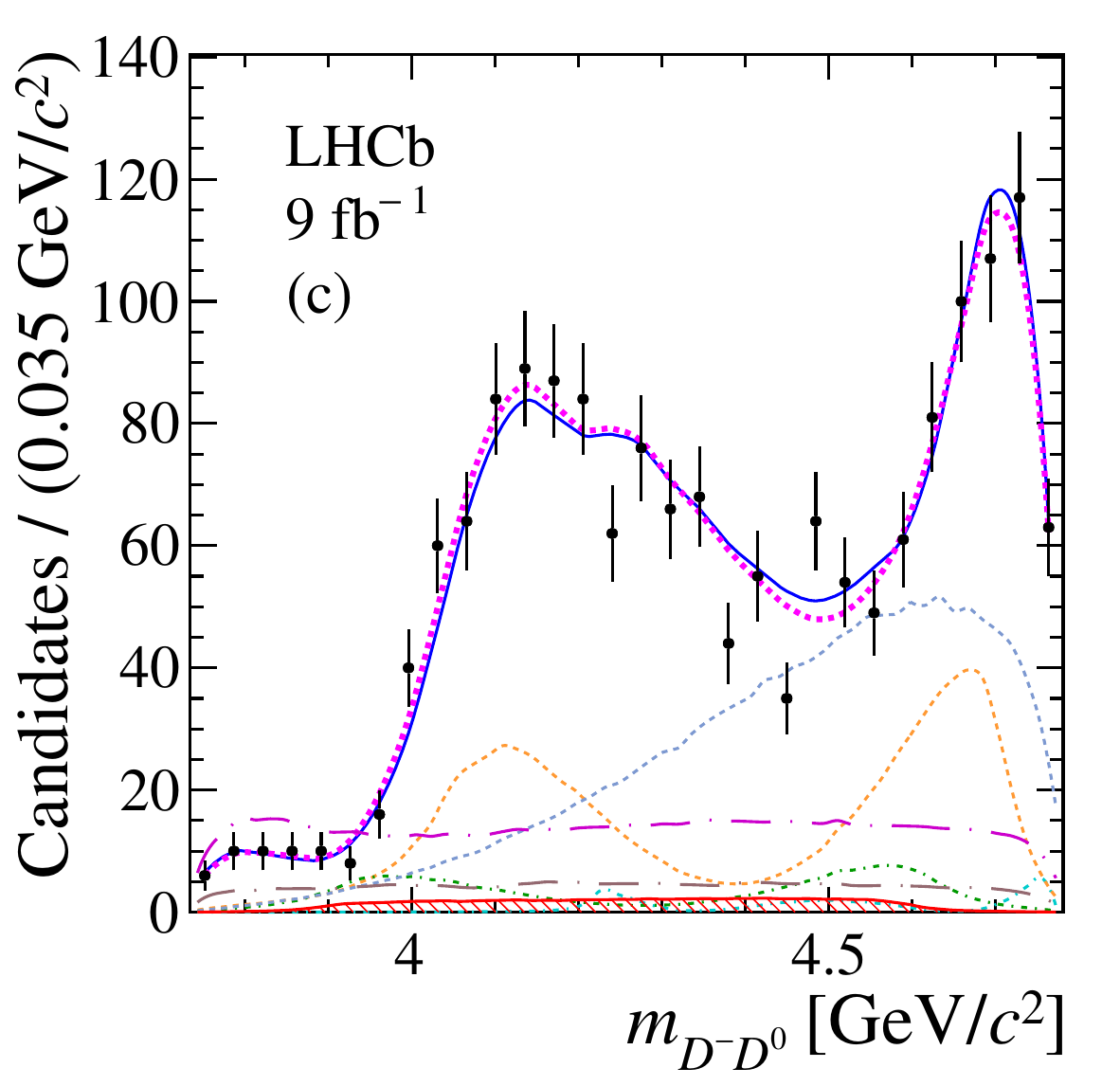}
    \caption{Mass distributions of  (a) $m_{D^- K_\mathrm{S}^0}$, (b) $m_{D^0 K_\mathrm{S}^0}$ and (c) $m_{D^- D^0}$, together with the fit projections with (thick blue) or without (dashed magenta) the $T_{cs0}^{*}(2870)^0$ state.
    The subcomponents correspond to the fit including the $T_{cs0}^{*}(2870)^0$ structure.
    }
    \label{fig:amp_fit}
\end{figure*}

The first model, used to fit the data, considers only resonant $\Dsstar$ mesons and nonresonant (NR) components in the $\Dm\KS$ channel. 
Resonances decaying strongly to two pseudoscalar mesons can only have spin-parity $J^P$ in the natural spin-parity series: $0^+$, $1^-$, $2^+$, \etc Those with masses above the $\Dm\KS$ mass threshold, $D_{s2}^*(2573)^-, D_{s1}^*(2700)^-$, and $D_{s1}^*(2860)^-$, are included in the fit.
The invariant-mass distribution  for each state is modeled using a relativistic Breit--Wigner (RBW) function, where the Blatt--Weisskopf radius is set to $4~(\!\gevc)^{-1}$~\cite{Back:2017zqt}, and the masses and widths  are fixed to their known values~\cite{PDG2024}. 
States with higher spin are less likely to be produced, and the effects of their possible presence are considered as a source of systematic uncertainty. 
In addition to $\Dsstar$ resonances, S-wave and P-wave NR contributions are needed to model the $m_{\Dm\KS}$ distribution, and described by an exponential and a uniform distribution, respectively. The slope of the exponential function is allowed to vary. The contribution of a D-wave NR component is included but found negligible and therefore not considered in the subsequent studies.
The two-body invariant-mass distributions and projections of the amplitude fit are shown in Fig.~\ref{fig:amp_fit}.
With these contributions from $\Dm\KS$  components,  the $m_{\Dm\KS}$ and $m_{\Dm\Dz}$ distributions are well described. 
However, the $m_{\Dz\KS}$ distribution is not well modeled, in particular around $2.9\gevcc$. 
Alternative fits varying the $\Dsstar$ model are performed, including varying masses and widths of $\Dsstar$ resonances within their uncertainties, changing the Blatt--Weisskopf radii, using the K-matrix parameterization~\cite{Chung:1995dx} to describe the spin-1 $\Dm\KS$ contributions, or adding higher spin $\Dm\KS$ resonances (\eg $D_{s3}^{*}(2860)^{-}$). 
None of the approaches significantly improve the fit quality around $m_{\Dz\KS}=2.9\gevcc$.
Charmoniumlike tetraquarks $T_{\cquark\cquarkbar}^-$ decaying into the $\Dm\Dz$ final state are also considered in the fit, but no significant signal is found.

The amplitude fit quality is improved by including a resonant contribution in the $\Dz\KS$ final state, with the invariant-mass distribution modeled by a RBW distribution with floated mass and width.
Various $J^P$ assignments are tested, with the $J^P=0^+$ giving the largest increase of the likelihood, with $2\Delta \ln\mathcal{L}=50.7$. This additional state is denoted as $\myTcsZeroShort$ in the following.
The mass and width of the $\myTcsZeroShort$ state are determined to be $2883\pm11\mevcc$ and $87_{-47}^{+22} \mev$, respectively, and the fit fraction (FF) of the \mbox{$\Bm\to \myTcsZeroShort(\to \Dz\KS)\Dm$} component in the total $\myBmDecay$ decay is $(2.6\pm1.2)\%$, where all uncertainties are statistical.  The uncertainty of the $\myTcsZeroShort$ width given by the fitter is corrected to be asymmetric through pseudoexperiments. The FF is calculated as the ratio between the phase-space integral of the squared amplitude for a single resonance to that of the total squared amplitude. The detailed results of the fit parameters are summarized in Table~\ref{tab:endmatter_amfitsummary} in the End Matter.
As shown in Fig.~\ref{fig:amp_fit}, for this configuration of the fit, all three invariant-mass distributions are well modeled by the fit. 
This fit is considered as the nominal model of the analysis.  
The significance of the $\myTcsZeroShort$ state is evaluated with a likelihood-ratio test using pseudoexperiments~\cite{LHCb-PAPER-2020-025}, to be $5.3$ standard deviations ($\sigma$) after accounting for systematic effects described later.
If instead  $J^P=1^-$ is set for the $\Dz\KS$ resonance (denoted as $\myTcsOneShort$),  its width is determined to be around $5\gev$ and the increase of the likelihood is only $2\Delta \ln\mathcal{L}=12.2$. 
A fit, including two $\Dz\KS$ states ($\ie$ $\myTcsZeroShort$ and $\myTcsOneShort$) with $J^P=0^+$ and $1^-$, respectively, has only a  marginal improvement compared to the fit including only the $\myTcsZeroShort$ state. Therefore, the contribution of any $\myTcsOneShort$ state is not significant, and its impact on the fit parameters is considered as a source of systematic uncertainty.

Various systematic uncertainties are considered for the mass, width and FF measurements of the $\myTcsZeroShort$ state, as summarized in Table~\ref{tab:syserr}. They are studied with alternative models, and largest deviations are taken as the systematic uncertainties. 
The uncertainty due to the $\Bm$ signal fraction $f_\text{s}$ is studied by varying $f_\text{s}$ within its uncertainty. 
The uncertainty due to the imperfect modeling of the background PDF in the amplitude fit is studied by using data samples with varied BDT selections 
to construct the distribution. 
The uncertainty due to the efficiency takes into account the contributions from
residual mismodeling of the $\Bm$ kinematics, PID and hardware-level trigger responses by simulation, studied using alternative data-driven corrections.
The uncertainty from the Blatt--Weisskopf radii are studied by varying from the default values of 4~$(\!\gevc)^{-1}$ to 3 or 5~$(\!\gevc)^{-1}$. 
The uncertainties due to known $\Dsstar$ masses and widths are studied by floating these parameters in the amplitude fit with Gaussian constraints to known values. 
Additional $D_{s3}^{*}(2860)^-$ and $\myTcsOne$ states are separately included in the amplitude fit to test the variation of the $\myTcsZeroShort$ measurements. 
The $\Dsstar$ model is also studied by using an alternative K-matrix model to describe the spin-1 $\Dm\KS$ contributions. 
The total systematic uncertainties are determined by combining all contributions in quadrature, and are found to be smaller than the statistical uncertainties.

\begin{table}[t]
    \centering
    \caption{Summary of systematic uncertainties on the mass, width and fit fraction of the $T_{cs0}^{*0}$ state observed in the $B^-\rightarrow D^- D^0 K_\mathrm{S}^0$ decay.  The total systematic uncertainties are determined by combining all contributions in quadrature.}
    \label{tab:syserr}
    \renewcommand{\arraystretch}{1.2}
    \begin{tabular}{lccc}
    \hline \hline 
Source  &Mass [$\mathrm{\!\,Me\kern -0.1em V\!/}c^2$]     &Width [$\mathrm{\!\,Me\kern -0.1em V}$] &FF [\%] \\
    \hline
    $f_\text{s}$ &0.9\phantom{0}  &\phantom{0}1.7   &0.06 \\
    Background PDF    &0.6\phantom{0}  &\phantom{0}2.1   &0.09 \\
    Efficiency &0.6\phantom{0} &\phantom{0}3.2    &0.11\\
    Blatt--Weisskopf radii   &1.2\phantom{0}   &\phantom{0}0.6  &0.02 \\
    $D_{sJ}^{*-}$ masses and widths  &4.0\phantom{0}   &\phantom{0}1.7   &0.01\\
    Fit with $D_{s3}^{*}(2860)^-$ &0.3\phantom{0}  &\phantom{0}2.5    &0.05   \\
    Fit with $\myTcsOne$ &4.2\phantom{0}   &15.7 &0.34 \\
    $D^-$--$K_\mathrm{S}^0$ K-matrix model     &5.0\phantom{0}   &\phantom{0}2.1   &0.04\\
    \hline
    Total &8\phantom{.00}  &17\phantom{.0}  &0.4\phantom{0}\\
    \hline\hline
    \end{tabular}
\end{table}

The mass, width and $J^P$ of the $\myTcsZeroShort$ state measured in the $\Dz\KS$ mass spectrum of the $\Bm\to\Dm\Dz\KS$ decay are consistent with those of the $\myTcsZero$ state observed in the $\Dp\Km$ mass spectrum of the $\Bm\to\Dm\Dp\Km$ decay~\cite{LHCb-PAPER-2020-025}, making them likely to be the same state.
Under this hypothesis, alternative fits are performed constraining the masses and widths of $\myTcsZeroOne$ resonances to the known values of  $\myTcsZero$ and  $\myTcsOne$ states~\cite{PDG2024}.
The fit fractions of the  $\myTcsZero$ and $\myTcsOne$ states  are measured to be
\mbox{$(2.6\pm0.9\pm0.5)\%$} and \mbox{$(0.6\pm0.6\pm0.2)\%$}, respectively, where the first uncertainties are statistical and the second systematic, and the systematic uncertainties include the sources listed in Table~\ref{tab:syserr}. 
The significance of the $\myTcsZero$ state is $6.4\,\sigma$, while the $\myTcsOne$ state alone has only a marginal significance of $1.8\,\sigma$, with effects of systematic uncertainties considered.

Isospin of genuine $\TcsGeneric$ states  in the $\PD\Kbar$ invariant-mass spectrum can be either $I=0$ or $1$. 
In either case, 
isospin symmetry requires the relative widths between the  $\TcsGeneric\to\Dz\KS$ and $\TcsGeneric\to\Dp\Km$  decays to be \mbox{$R_\text{I}(\TcsGeneric)\equiv\Gamma(\TcsGeneric\to\Dz\Kzb)/\Gamma(\TcsGeneric\to\Dp\Km)\approx 1$}. In addition, the ratio between the fit fractions of the $\myTcsOneShort$ and $\myTcsZeroShort$ states in the $\myBmDecay$ decay, 
\begin{equation}
    R_\text{FF}(\Dz\Kzb)\equiv\frac{\text{FF}(\myTcsOneShort\to \Dz\Kzb)}{\text{FF}(\myTcsZeroShort\to\Dz\Kzb)}=\frac{\text{FF}(\myTcsOneShort\to \Dz\KS)}{\text{FF}(\myTcsZeroShort\to\Dz\KS)},
\end{equation}
is expected to approximate that in the $\Bm\to\Dm\Dp\Km$ decay, $R_\text{FF}(\Dp\Km)$. Here, $\text{FF}(T_{cs}^{*0}\to\Dz\Kzb)$ is the fit fraction of the $T_{cs}^{*0}\to \Dz\Kzb$ components in the $\Bm\to\Dm\Dz\Kzb$ decay and is equal to $\text{FF}(\TcsGeneric\to\Dz\KS)$.
The relative decay widths $R_\text{I}(\TcsGeneric)$ and the double ratio \mbox{$ R_\text{FF}(\Dz\Kzb)/R_\text{FF}(\Dp\Km)$} provide precise tests of the isospin symmetry. The relative widths $R_\text{I}(\TcsGeneric)$ are calculated according to
\begin{equation}
    R_\text{I}(\TcsGeneric)=\frac{\BR(\Bm\to\Dm\Dz\Kzb)\text{FF}(\TcsGeneric\to\Dz\KS)}{\BR(\Bm\to\Dm\Dp\Km)\text{FF}(\TcsGeneric\to\Dp\Km)},\label{eq:RI}
\end{equation}
where $\BR(\Bm\to\Dm\Dz\Kzb)$ and $\BR(\Bm\to\Dm\Dp\Km)$  are the respective branching fractions~\cite{PDG2024}.
The ratio of \mbox{$R_\text{FF}(\Dz\KS)/R_\text{FF}(\Dp\Km)$} is equivalent to that of the relative decay widths as
\begin{equation}
    \frac{R_\text{FF}(\Dz\Kzb)}{R_\text{FF}(\Dp\Km)}=\frac{R_\text{I}[\myTcsOne]}{R_\text{I}[\myTcsZero]}.
    \label{eq:RFF/RFF}
\end{equation}

The quantities $R_\text{I}(\TcsGeneric)$ and $R_\text{FF}(\Dz\Kzb)/R_\text{FF}(\Dp\Km)$ are measured assuming the $\myTcsZeroOne$ states observed in the $\decay{\Bm}{\Dm\Dp\Km}$ decay are also present in the $\Bm\to\Dm\Dz\KS$ decay.
As shown in Table~\ref{tab:isospin_sym},  the relative rate for the $\myTcsZero$ decays, $R_\text{I}[\myTcsZero]$, is  consistent with the hypothesis of isospin invariance of a genuine state, although a fairly large isospin violation cannot be excluded at the current precision.
However, the relative rate for the $\myTcsOne$ decays $R_\text{I}[\myTcsOne]$  and the double fit fraction ratio $R_\text{FF}(\Dz\Kzb)/R_\text{FF}(\Dp\Km)$ are significantly smaller than unity, which indicates isospin violation between $\myTcsOne$ decays. 
The violation can be explained if the $\myTcsOne$ structure is caused by  kinematic singularities of  $\Bm\to D^{(*)0}\D_{sJ}^{(*)-}(\to \Dm K^{(*)0})$ decays with final-state rescattering or if the $\myTcsOne$ structure as a genuine state does not have a definite isospin~\cite{Burns:2020xne}.

\begin{table}[t]
    \centering
    \caption{
    Measurements of the relative rates of decays into the $D^0  \kern 0.18em\overline{\kern -0.18em K}^{0} $ and $D^+ K^-$ final states for $T_{cs0}^*(2870)^0$ and $T_{cs1}^*(2900)^0$ states, $R_\text{I}[T_{cs0}^*(2870)^0]$ and $R_\text{I}[T_{cs1}^*(2900)^0]$, and the double fit fraction ratio $R_\text{FF}(\Dz\Kzb)/R_\text{FF}(\Dp\Km)$.
    The first, second and third uncertainties are statistical, systematic, and due to the external inputs of $B^-\rightarrow D^- D^+ K^-$ and $B^- \rightarrow D^- D^0 K_\mathrm{S}^0$ branching fractions, respectively.}
    
    \label{tab:isospin_sym}
    \renewcommand{\arraystretch}{1.8}
    \begin{tabular}{lc}
    \hline \hline
    Observable    &Result \\
        \hline
   $R_\text{I}[T_{cs0}^*(2870)^0]$        &$3.3\phantom{0}\phantom{0}\pm1.1\phantom{0}\phantom{0}\pm1.1\phantom{0}\phantom{0}\pm1.1\phantom{0}\phantom{0}$  \\
   $R_\text{I}[T_{cs1}^*(2900)^0]$        &$0.15\phantom{0}\pm0.15\phantom{0}\pm0.05\phantom{0}\pm0.05\phantom{0}$  \\
  $R_\text{FF}(\Dz\Kzb)/R_\text{FF}(\Dp\Km)$   &$0.044\pm0.035\pm0.020\phantom{0000000}$ \\
   \hline\hline
    \end{tabular}
\end{table}

In conclusion, an amplitude analysis of the $\myBmDecay$ decay is performed, using $pp$ collisions at center-of-mass energies $7, 8$ and $13\tev$ collected by the \lhcb detector, corresponding to an integrated luminosity of $9\invfb$. 
A spin-0 open-charm tetraquark candidate $\myTcsZeroShort$ is observed in the $\Dz\KS$ final state for the first time, with a significance of $5.3\,\sigma$.
The mass and width of this state and the fit fraction in the \myBmDecay decay are  measured to be
\begin{align*}
        M(\myTcsZeroShort)&=2883\pm11\pm8 \mevcc, \\
        \Gamma(\myTcsZeroShort)&=87_{-47}^{+22}\pm17 \mev, \\
        \text{FF}(\myTcsZeroShort\to\Dz\KS)&=(2.6\pm1.2\pm0.4)\%,
\end{align*}
where the first uncertainties are statistical and the second systematic. The mass, width, spin-parity and flavor content are all consistent with those of the $\myTcsZero$ state observed in the $\Dp\Km$ invariant-mass spectrum of the $\Bm\to\Dm\Dp\Km$ decay~\cite{LHCb-PAPER-2020-025}.
No significant $\TcsGeneric$ states with $J^P=1^-$ or charmoniumlike tetraquarks are observed in the $\myBmDecay$ decay.
With masses and widths constrained to known values, the $\myTcsZero$ state is observed with a significance of $6.4\,\sigma$ in the $\myBmDecay$ decay, while the $\myTcsOne$ state alone has a marginal significance of $1.8\,\sigma$.  
Assuming $\myTcsZero$ and $\myTcsOne$ states to be genuine hadrons, a violation of isospin invariance between the $\myTcsOne\to \Dz\Kzb$ and the $\myTcsOne\to\Dp\Km$ decay rates is indicated. On the other hand, the relative rate between the $\myTcsZero\to \Dz\Kzb$ and $\myTcsZero\to \Dp\Km$ decays is consistent with an isospin invariance at current experimental precision.
The study in this analysis helps to shed light on the nature of the $\myTcsZeroOne$ states, in particular on the resonance or kinematic singularity interpretations.

\section*{Acknowledgements}
%
%
\noindent We express our gratitude to our colleagues in the CERN
accelerator departments for the excellent performance of the LHC. We
thank the technical and administrative staff at the LHCb
institutes.
We acknowledge support from CERN and from the national agencies:
CAPES, CNPq, FAPERJ and FINEP (Brazil); 
MOST and NSFC (China); 
CNRS/IN2P3 (France); 
BMBF, DFG and MPG (Germany); 
INFN (Italy); 
NWO (Netherlands); 
MNiSW and NCN (Poland); 
MCID/IFA (Romania); 
MICIU and AEI (Spain);
SNSF and SER (Switzerland); 
NASU (Ukraine); 
STFC (United Kingdom); 
DOE NP and NSF (USA).
We acknowledge the computing resources that are provided by CERN, IN2P3
(France), KIT and DESY (Germany), INFN (Italy), SURF (Netherlands),
PIC (Spain), GridPP (United Kingdom), 
CSCS (Switzerland), IFIN-HH (Romania), CBPF (Brazil),
and Polish WLCG (Poland).
We are indebted to the communities behind the multiple open-source
software packages on which we depend.
Individual groups or members have received support from
ARC and ARDC (Australia);
Key Research Program of Frontier Sciences of CAS, CAS PIFI, CAS CCEPP, 
Fundamental Research Funds for the Central Universities, 
and Sci. \& Tech. Program of Guangzhou (China);
Minciencias (Colombia);
EPLANET, Marie Sk\l{}odowska-Curie Actions, ERC and NextGenerationEU (European Union);
A*MIDEX, ANR, IPhU and Labex P2IO, and R\'{e}gion Auvergne-Rh\^{o}ne-Alpes (France);
AvH Foundation (Germany);
ICSC (Italy); 
Severo Ochoa and Mar\'ia de Maeztu Units of Excellence, GVA, XuntaGal, GENCAT, InTalent-Inditex and Prog. ~Atracci\'on Talento CM (Spain);
SRC (Sweden);
the Leverhulme Trust, the Royal Society
 and UKRI (United Kingdom).

\clearpage
\section*{End Matter}
\textit{Amplitude fit results using the nominal model} — The nominal amplitude fit results are summarized  in Table~\ref{tab:endmatter_amfitsummary}.

\begin{table}[!hbtp]
    \centering
    \caption{Fit results of the parameters in the amplitude fit using the nominal amplitude model. The first uncertainty is statistical, and the second is systematic. The \Real, \Imag, FF represent the real and imaginary part of the helicity couplings and the fit fraction respectively.}
    \label{tab:endmatter_amfitsummary}
    \begin{tabular}{ccc}
        \hline\hline
        Decay channel   & Parameter  &Fit result\\
        \hline
        \multirow{3}{*}{$\decay{\Bm}{D_{s1}^{*}(2700)^-\Dz}$} &\Real &1 (fixed) \\
            &\Imag&0 (fixed)  \\
            &FF   &$\phantom{-}0.271\pm0.029\pm0.020$\\
        \hline
        \multirow{3}{*}{$\decay{\Bm}{D_{s2}^{*}(2573)^-\Dz}$} &\Real &$-0.043\pm0.019\pm0.042$\\
        &\Imag&$-0.109\pm0.019\pm0.040$\\
        &FF   &$0.0159\pm0.0047\pm0.0020$\\
        \hline
        \multirow{3}{*}{$\decay{\Bm}{D_{s1}^{*}(2860)^-\Dz}$} &\Real &$-0.30\pm0.10\pm0.06$\\
            &\Imag&$\phantom{-}0.45\pm0.10\pm0.18$\\
            &FF   &$0.060\pm0.015\pm0.010$\\
        \hline
            \multirow{4}{*}{$\decay{\Bm}{\text{NR (S-wave) }\Dz}$}  &\Real &$\phantom{-}1.17\pm0.22\pm0.30$\\
                &\Imag&$-2.18\pm0.17\pm1.43$\\
                &Slope  &$0.476\pm0.033\pm0.018~ (\!\gevcc)^{-2}$\\
                &FF   &$0.449\pm0.030\pm0.038$\\
                \hline
            \multirow{3}{*}{$\decay{\Bm}{\text{NR (P-wave) }\Dz}$} &\Real  &$1.44\pm0.15\pm0.18$\\
                &\Imag&$0.21\pm0.17\pm0.11$\\
                &FF   &$0.271\pm0.034\pm0.025$\\
            \hline
            \multirow{5}{*}{$\decay{\Bm}{\myTcsZero\Dm}$}    &\Real &$-0.12\pm0.04\pm0.06$\\
                &\Imag&$\phantom{-}0.09\pm0.05\pm0.09$\\
                &FF   &$0.0259\pm0.012\pm0.004$\\
                &Mass  &$2883\pm11\pm8\mevcc$\\
                &Width  &$\phantom{00}87\,\, _{-\,\,\,\,47}^{+\,\,\,\,22}\,\pm17\mev\phantom{\!/c^2}$\\
            \hline
            Total fit fraction  &   &$1.093\pm0.026$ \\
        \hline\hline
    
    \end{tabular}
\end{table}

\clearpage
\addcontentsline{toc}{section}{References}
\bibliographystyle{LHCb}
\bibliography{main,standard,LHCb-PAPER,LHCb-CONF,LHCb-DP,LHCb-TDR}

\ifx\mcitethebibliography\mciteundefinedmacro
\PackageError{LHCb.bst}{mciteplus.sty has not been loaded}
{This bibstyle requires the use of the mciteplus package.}\fi
\providecommand{\href}[2]{#2}
\begin{mcitethebibliography}{10}
\mciteSetBstSublistMode{n}
\mciteSetBstMaxWidthForm{subitem}{\alph{mcitesubitemcount})}
\mciteSetBstSublistLabelBeginEnd{\mcitemaxwidthsubitemform\space}
{\relax}{\relax}

\bibitem{Liu:2019zoy}
Y.-R. Liu {\em et~al.}, \ifthenelse{\boolean{articletitles}}{\emph{{Pentaquark
  and tetraquark states}},
  }{}\href{https://doi.org/10.1016/j.ppnp.2019.04.003}{Prog.\ Part.\ Nucl.\
  Phys.\  \textbf{107} (2019) 237},
  \href{http://arxiv.org/abs/1903.11976}{{\normalfont\ttfamily
  arXiv:1903.11976}}\relax
\mciteBstWouldAddEndPuncttrue
\mciteSetBstMidEndSepPunct{\mcitedefaultmidpunct}
{\mcitedefaultendpunct}{\mcitedefaultseppunct}\relax
\EndOfBibitem
\bibitem{Guo:2017jvc}
F.-K. Guo {\em et~al.}, \ifthenelse{\boolean{articletitles}}{\emph{Hadronic
  molecules}, }{}\href{https://doi.org/10.1103/RevModPhys.90.015004}{Rev.\
  Mod.\ Phys.\  \textbf{90} (2018) 015004},
  \href{http://arxiv.org/abs/1705.00141}{{\normalfont\ttfamily
  arXiv:1705.00141}}\relax
\mciteBstWouldAddEndPuncttrue
\mciteSetBstMidEndSepPunct{\mcitedefaultmidpunct}
{\mcitedefaultendpunct}{\mcitedefaultseppunct}\relax
\EndOfBibitem
\bibitem{Belle:2003nnu}
Belle collaboration, S.-K. Choi {\em et~al.},
  \ifthenelse{\boolean{articletitles}}{\emph{{Observation of a narrow
  charmoniumlike state in exclusive $B^\pm \to K^\pm \pi^+ \pi^- \jpsi$
  decays}}, }{}\href{https://doi.org/10.1103/PhysRevLett.91.262001}{Phys.\
  Rev.\ Lett.\  \textbf{91} (2003) 262001},
  \href{http://arxiv.org/abs/hep-ex/0309032}{{\normalfont\ttfamily
  arXiv:hep-ex/0309032}}\relax
\mciteBstWouldAddEndPuncttrue
\mciteSetBstMidEndSepPunct{\mcitedefaultmidpunct}
{\mcitedefaultendpunct}{\mcitedefaultseppunct}\relax
\EndOfBibitem
\bibitem{BESIII:2013ris}
BESIII collaboration, M.~Ablikim {\em et~al.},
  \ifthenelse{\boolean{articletitles}}{\emph{{Observation of a charged
  charmoniumlike structure in $e^+e^- \to \pi^+\pi^- \jpsi$ at $\sqrt{s}
  =4.26\gev$}}, }{}\href{https://doi.org/10.1103/PhysRevLett.110.252001}{Phys.\
  Rev.\ Lett.\  \textbf{110} (2013) 252001},
  \href{http://arxiv.org/abs/1303.5949}{{\normalfont\ttfamily
  arXiv:1303.5949}}\relax
\mciteBstWouldAddEndPuncttrue
\mciteSetBstMidEndSepPunct{\mcitedefaultmidpunct}
{\mcitedefaultendpunct}{\mcitedefaultseppunct}\relax
\EndOfBibitem
\bibitem{Belle:2013yex}
{Belle} collaboration, Z.~Q. Liu {\em et~al.},
  \ifthenelse{\boolean{articletitles}}{\emph{{Study of $e^+e^-\to\pi^+\pi^-
  \jpsi$ and observation of a charged charmoniumlike state at Belle}},
  }{}\href{https://doi.org/10.1103/PhysRevLett.110.252002}{{Phys.\ Rev.\ Lett.\
  } \textbf{{110}} (2013) 252002},
  \href{http://arxiv.org/abs/1304.0121}{{\normalfont\ttfamily
  arXiv:1304.0121}}\relax
\mciteBstWouldAddEndPuncttrue
\mciteSetBstMidEndSepPunct{\mcitedefaultmidpunct}
{\mcitedefaultendpunct}{\mcitedefaultseppunct}\relax
\EndOfBibitem
\bibitem{LHCb-PAPER-2021-031}
LHCb collaboration, R.~Aaij {\em et~al.},
  \ifthenelse{\boolean{articletitles}}{\emph{{Observation of an exotic narrow
  doubly charmed tetraquark}},
  }{}\href{https://doi.org/10.1038/s41567-022-01614-y}{Nature Physics
  \textbf{18} (2022) 751},
  \href{http://arxiv.org/abs/2109.01038}{{\normalfont\ttfamily
  arXiv:2109.01038}}\relax
\mciteBstWouldAddEndPuncttrue
\mciteSetBstMidEndSepPunct{\mcitedefaultmidpunct}
{\mcitedefaultendpunct}{\mcitedefaultseppunct}\relax
\EndOfBibitem
\bibitem{LHCb-PAPER-2021-032}
LHCb collaboration, R.~Aaij {\em et~al.},
  \ifthenelse{\boolean{articletitles}}{\emph{{Study of the doubly charmed
  tetraquark $T^+_{cc}$}},
  }{}\href{https://doi.org/10.1038/s41467-022-30206-w}{Nature Communications
  \textbf{13} (2022) 3351},
  \href{http://arxiv.org/abs/2109.01056}{{\normalfont\ttfamily
  arXiv:2109.01056}}\relax
\mciteBstWouldAddEndPuncttrue
\mciteSetBstMidEndSepPunct{\mcitedefaultmidpunct}
{\mcitedefaultendpunct}{\mcitedefaultseppunct}\relax
\EndOfBibitem
\bibitem{LHCb-PAPER-2020-011}
LHCb collaboration, R.~Aaij {\em et~al.},
  \ifthenelse{\boolean{articletitles}}{\emph{{Observation of structure in the
  \jpsi-pair mass spectrum}},
  }{}\href{https://doi.org/10.1016/j.scib.2020.08.032}{Science Bulletin
  \textbf{65} (2020) 1983},
  \href{http://arxiv.org/abs/2006.16957}{{\normalfont\ttfamily
  arXiv:2006.16957}}\relax
\mciteBstWouldAddEndPuncttrue
\mciteSetBstMidEndSepPunct{\mcitedefaultmidpunct}
{\mcitedefaultendpunct}{\mcitedefaultseppunct}\relax
\EndOfBibitem
\bibitem{ATLAS:2023bft}
ATLAS collaboration, G.~Aad {\em et~al.},
  \ifthenelse{\boolean{articletitles}}{\emph{{Observation of an excess of
  dicharmonium events in the four-muon final state with the ATLAS detector}},
  }{}\href{https://doi.org/10.1103/PhysRevLett.131.151902}{Phys.\ Rev.\ Lett.\
  \textbf{131} (2023) 151902},
  \href{http://arxiv.org/abs/2304.08962}{{\normalfont\ttfamily
  arXiv:2304.08962}}\relax
\mciteBstWouldAddEndPuncttrue
\mciteSetBstMidEndSepPunct{\mcitedefaultmidpunct}
{\mcitedefaultendpunct}{\mcitedefaultseppunct}\relax
\EndOfBibitem
\bibitem{CMS:2023owd}
CMS collaboration, A.~Hayrapetyan {\em et~al.},
  \ifthenelse{\boolean{articletitles}}{\emph{{New structures in the $\jpsi
  \jpsi$ mass spectrum in proton-proton collisions at $\sqrt{s}=13\tev$}},
  }{}\href{https://doi.org/10.1103/PhysRevLett.132.111901}{Phys.\ Rev.\ Lett.\
  \textbf{132} (2024) 111901},
  \href{http://arxiv.org/abs/2306.07164}{{\normalfont\ttfamily
  arXiv:2306.07164}}\relax
\mciteBstWouldAddEndPuncttrue
\mciteSetBstMidEndSepPunct{\mcitedefaultmidpunct}
{\mcitedefaultendpunct}{\mcitedefaultseppunct}\relax
\EndOfBibitem
\bibitem{LHCb-PAPER-2015-029}
LHCb collaboration, R.~Aaij {\em et~al.},
  \ifthenelse{\boolean{articletitles}}{\emph{{Observation of $\jpsi\proton$
  resonances consistent with pentaquark states in
  \mbox{\decay{\Lb}{\jpsi\proton\Km}} decays}},
  }{}\href{https://doi.org/10.1103/PhysRevLett.115.072001}{Phys.\ Rev.\ Lett.\
  \textbf{115} (2015) 072001},
  \href{http://arxiv.org/abs/1507.03414}{{\normalfont\ttfamily
  arXiv:1507.03414}}\relax
\mciteBstWouldAddEndPuncttrue
\mciteSetBstMidEndSepPunct{\mcitedefaultmidpunct}
{\mcitedefaultendpunct}{\mcitedefaultseppunct}\relax
\EndOfBibitem
\bibitem{LHCb-PAPER-2019-014}
LHCb collaboration, R.~Aaij {\em et~al.},
  \ifthenelse{\boolean{articletitles}}{\emph{{Observation of a narrow
  pentaquark state, $P_c(4312)^+$, and of two-peak structure of the
  $P_c(4450)^+$}},
  }{}\href{https://doi.org/10.1103/PhysRevLett.122.222001}{Phys.\ Rev.\ Lett.\
  \textbf{122} (2019) 222001},
  \href{http://arxiv.org/abs/1904.03947}{{\normalfont\ttfamily
  arXiv:1904.03947}}\relax
\mciteBstWouldAddEndPuncttrue
\mciteSetBstMidEndSepPunct{\mcitedefaultmidpunct}
{\mcitedefaultendpunct}{\mcitedefaultseppunct}\relax
\EndOfBibitem
\bibitem{LHCb-PAPER-2021-018}
LHCb collaboration, R.~Aaij {\em et~al.},
  \ifthenelse{\boolean{articletitles}}{\emph{{Evidence for a new structure in
  the $\jpsi \proton$ and $\jpsi \antiproton$ systems in $\Bs \to
  \jpsi\proton\antiproton$ decays}},
  }{}\href{https://doi.org/10.1103/PhysRevLett.128.062001}{Phys.\ Rev.\ Lett.\
  \textbf{128} (2022) 062001},
  \href{http://arxiv.org/abs/2108.04720}{{\normalfont\ttfamily
  arXiv:2108.04720}}\relax
\mciteBstWouldAddEndPuncttrue
\mciteSetBstMidEndSepPunct{\mcitedefaultmidpunct}
{\mcitedefaultendpunct}{\mcitedefaultseppunct}\relax
\EndOfBibitem
\bibitem{LHCb-PAPER-2020-039}
LHCb collaboration, R.~Aaij {\em et~al.},
  \ifthenelse{\boolean{articletitles}}{\emph{{Evidence of a $\jpsi\Lz$
  structure and observation of excited $\Xim$ states in the $\Xibm \rightarrow
  \jpsi \Lz \Km$ decay}},
  }{}\href{https://doi.org/10.1016/j.scib.2021.02.030}{Science Bulletin
  \textbf{66} (2021) 1278},
  \href{http://arxiv.org/abs/2012.10380}{{\normalfont\ttfamily
  arXiv:2012.10380}}\relax
\mciteBstWouldAddEndPuncttrue
\mciteSetBstMidEndSepPunct{\mcitedefaultmidpunct}
{\mcitedefaultendpunct}{\mcitedefaultseppunct}\relax
\EndOfBibitem
\bibitem{LHCb-PAPER-2022-031}
LHCb collaboration, R.~Aaij {\em et~al.},
  \ifthenelse{\boolean{articletitles}}{\emph{{Observation of a
  $\jpsi\Lambdares$ resonance consistent with a strange pentaquark candidate in
  $\Bm\to \jpsi\Lambdares\antiproton$ decays}},
  }{}\href{https://doi.org/10.1103/PhysRevLett.131.031901}{Phys.\ Rev.\ Lett.\
  \textbf{131} (2023) 031901},
  \href{http://arxiv.org/abs/2210.10346}{{\normalfont\ttfamily
  arXiv:2210.10346}}\relax
\mciteBstWouldAddEndPuncttrue
\mciteSetBstMidEndSepPunct{\mcitedefaultmidpunct}
{\mcitedefaultendpunct}{\mcitedefaultseppunct}\relax
\EndOfBibitem
\bibitem{LHCb-PAPER-2020-025}
LHCb collaboration, R.~Aaij {\em et~al.},
  \ifthenelse{\boolean{articletitles}}{\emph{{Amplitude analysis of the $\Bp
  \to \Dp \Dm \Kp$ decay}},
  }{}\href{https://doi.org/10.1103/PhysRevD.102.112003}{Phys.\ Rev.\
  \textbf{D102} (2020) 112003},
  \href{http://arxiv.org/abs/2009.00026}{{\normalfont\ttfamily
  arXiv:2009.00026}}\relax
\mciteBstWouldAddEndPuncttrue
\mciteSetBstMidEndSepPunct{\mcitedefaultmidpunct}
{\mcitedefaultendpunct}{\mcitedefaultseppunct}\relax
\EndOfBibitem
\bibitem{LHCb-PAPER-2020-024}
LHCb collaboration, R.~Aaij {\em et~al.},
  \ifthenelse{\boolean{articletitles}}{\emph{{Model-independent study of
  structure in \mbox{$\Bp \to \Dp \Dm \Kp$} decays}},
  }{}\href{https://doi.org/10.1103/PhysRevLett.125.242001}{Phys.\ Rev.\ Lett.\
  \textbf{125} (2020) 242001},
  \href{http://arxiv.org/abs/2009.00025}{{\normalfont\ttfamily
  arXiv:2009.00025}}\relax
\mciteBstWouldAddEndPuncttrue
\mciteSetBstMidEndSepPunct{\mcitedefaultmidpunct}
{\mcitedefaultendpunct}{\mcitedefaultseppunct}\relax
\EndOfBibitem
\bibitem{LHCb-PAPER-2023-047}
LHCb collaboration, R.~Aaij {\em et~al.},
  \ifthenelse{\boolean{articletitles}}{\emph{{Observation of new charmonium or
  charmoniumlike states in $B^+\to D^{*\pm} D^{\mp} K^+$ decays}},
  }{}\href{https://doi.org/10.1103/PhysRevLett.133.131902}{{Phys.\ Rev.\ Lett.\
  } \textbf{133} (2024) 131902},
  \href{http://arxiv.org/abs/2406.03156}{{\normalfont\ttfamily
  arXiv:2406.03156}}\relax
\mciteBstWouldAddEndPuncttrue
\mciteSetBstMidEndSepPunct{\mcitedefaultmidpunct}
{\mcitedefaultendpunct}{\mcitedefaultseppunct}\relax
\EndOfBibitem
\bibitem{LHCb-PAPER-2023-042}
LHCb collaboration, R.~Aaij {\em et~al.},
  \ifthenelse{\boolean{articletitles}}{\emph{{First observation of the $\Lb \to
  \Dp\Dm \Lz$ decay}}, }{}\href{https://doi.org/10.1007/JHEP07(2024)140}{{JHEP}
  \textbf{07} (2024) 140},
  \href{http://arxiv.org/abs/2403.03586}{{\normalfont\ttfamily
  arXiv:2403.03586}}\relax
\mciteBstWouldAddEndPuncttrue
\mciteSetBstMidEndSepPunct{\mcitedefaultmidpunct}
{\mcitedefaultendpunct}{\mcitedefaultseppunct}\relax
\EndOfBibitem
\bibitem{LHCb-PAPER-2022-026}
LHCb collaboration, R.~Aaij {\em et~al.},
  \ifthenelse{\boolean{articletitles}}{\emph{{First observation of a doubly
  charged tetraquark candidate and its neutral partner}},
  }{}\href{https://doi.org/10.1103/PhysRevLett.131.041902}{Phys.\ Rev.\ Lett.\
  \textbf{131} (2023) 041902},
  \href{http://arxiv.org/abs/2212.02716}{{\normalfont\ttfamily
  arXiv:2212.02716}}\relax
\mciteBstWouldAddEndPuncttrue
\mciteSetBstMidEndSepPunct{\mcitedefaultmidpunct}
{\mcitedefaultendpunct}{\mcitedefaultseppunct}\relax
\EndOfBibitem
\bibitem{LHCb-PAPER-2022-027}
LHCb collaboration, R.~Aaij {\em et~al.},
  \ifthenelse{\boolean{articletitles}}{\emph{{Amplitude analysis of $\Bz
  \rightarrow \Dzb \Dsp \pim$ and $\Bp \rightarrow \Dm \Dsp\pip$ decays}},
  }{}\href{https://doi.org/10.1103/PhysRevD.108.012017}{Phys.\ Rev.\
  \textbf{D108} (2023) 012017},
  \href{http://arxiv.org/abs/2212.02717}{{\normalfont\ttfamily
  arXiv:2212.02717}}\relax
\mciteBstWouldAddEndPuncttrue
\mciteSetBstMidEndSepPunct{\mcitedefaultmidpunct}
{\mcitedefaultendpunct}{\mcitedefaultseppunct}\relax
\EndOfBibitem
\bibitem{Karliner:2020vsi}
M.~Karliner and J.~L. Rosner, \ifthenelse{\boolean{articletitles}}{\emph{First
  exotic hadron with open heavy flavor: $\cquark\squark\uquarkbar\dquarkbar$
  tetraquark}, }{}\href{https://doi.org/10.1103/PhysRevD.102.094016}{Phys.\
  Rev.\  \textbf{D102} (2020) 094016},
  \href{http://arxiv.org/abs/2008.05993}{{\normalfont\ttfamily
  arXiv:2008.05993}}\relax
\mciteBstWouldAddEndPuncttrue
\mciteSetBstMidEndSepPunct{\mcitedefaultmidpunct}
{\mcitedefaultendpunct}{\mcitedefaultseppunct}\relax
\EndOfBibitem
\bibitem{He:2020jna}
X.-G. He, W.~Wang, and R.~Zhu,
  \ifthenelse{\boolean{articletitles}}{\emph{{Open-charm tetraquark $X_c$ and
  open-bottom tetraquark $X_b$}},
  }{}\href{https://doi.org/10.1140/epjc/s10052-020-08597-1}{Eur.\ Phys.\ J.\
  \textbf{C80} (2020) 1026},
  \href{http://arxiv.org/abs/2008.07145}{{\normalfont\ttfamily
  arXiv:2008.07145}}\relax
\mciteBstWouldAddEndPuncttrue
\mciteSetBstMidEndSepPunct{\mcitedefaultmidpunct}
{\mcitedefaultendpunct}{\mcitedefaultseppunct}\relax
\EndOfBibitem
\bibitem{Guo:2021mja}
T.~Guo, J.~Li, J.~Zhao, and L.~He,
  \ifthenelse{\boolean{articletitles}}{\emph{{Mass spectra and decays of
  open-heavy tetraquark states}},
  }{}\href{https://doi.org/10.1103/PhysRevD.105.054018}{Phys.\ Rev.\
  \textbf{D105} (2022) 054018},
  \href{http://arxiv.org/abs/2108.06222}{{\normalfont\ttfamily
  arXiv:2108.06222}}\relax
\mciteBstWouldAddEndPuncttrue
\mciteSetBstMidEndSepPunct{\mcitedefaultmidpunct}
{\mcitedefaultendpunct}{\mcitedefaultseppunct}\relax
\EndOfBibitem
\bibitem{Ozdem:2022ydv}
U.~\"Ozdem and K.~Azizi, \ifthenelse{\boolean{articletitles}}{\emph{{Magnetic
  moment of the $X_1(2900)$ state in the diquark\textendash{}antidiquark
  picture}}, }{}\href{https://doi.org/10.1140/epja/s10050-022-00815-6}{Eur.\
  Phys.\ J.\  \textbf{A58} (2022) 171},
  \href{http://arxiv.org/abs/2202.11466}{{\normalfont\ttfamily
  arXiv:2202.11466}}\relax
\mciteBstWouldAddEndPuncttrue
\mciteSetBstMidEndSepPunct{\mcitedefaultmidpunct}
{\mcitedefaultendpunct}{\mcitedefaultseppunct}\relax
\EndOfBibitem
\bibitem{Wang:2020xyc}
Z.-G. Wang, \ifthenelse{\boolean{articletitles}}{\emph{{Analysis of the
  $X_0(2900)$ as the scalar tetraquark state via the QCD sum rules}},
  }{}\href{https://doi.org/10.1142/S0217751X20501870}{Int.\ J.\ Mod.\ Phys.\
  \textbf{A35} (2020) 2050187},
  \href{http://arxiv.org/abs/2008.07833}{{\normalfont\ttfamily
  arXiv:2008.07833}}\relax
\mciteBstWouldAddEndPuncttrue
\mciteSetBstMidEndSepPunct{\mcitedefaultmidpunct}
{\mcitedefaultendpunct}{\mcitedefaultseppunct}\relax
\EndOfBibitem
\bibitem{Agaev:2021knl}
S.~S. Agaev, K.~Azizi, and H.~Sundu,
  \ifthenelse{\boolean{articletitles}}{\emph{{Vector resonance $X_1(2900)$ and
  its structure}},
  }{}\href{https://doi.org/10.1016/j.nuclphysa.2021.122202}{Nucl.\ Phys.\
  \textbf{A1011} (2021) 122202},
  \href{http://arxiv.org/abs/2103.06151}{{\normalfont\ttfamily
  arXiv:2103.06151}}\relax
\mciteBstWouldAddEndPuncttrue
\mciteSetBstMidEndSepPunct{\mcitedefaultmidpunct}
{\mcitedefaultendpunct}{\mcitedefaultseppunct}\relax
\EndOfBibitem
\bibitem{Yang:2021izl}
G.~Yang, J.~Ping, and J.~Segovia,
  \ifthenelse{\boolean{articletitles}}{\emph{{$sQ\quarkbar\quarkbar$
  $(q=u,\,d;\, Q=c,\,b)$ tetraquarks in the chiral quark model}},
  }{}\href{https://doi.org/10.1103/PhysRevD.103.074011}{Phys.\ Rev.\
  \textbf{D103} (2021) 074011},
  \href{http://arxiv.org/abs/2101.04933}{{\normalfont\ttfamily
  arXiv:2101.04933}}\relax
\mciteBstWouldAddEndPuncttrue
\mciteSetBstMidEndSepPunct{\mcitedefaultmidpunct}
{\mcitedefaultendpunct}{\mcitedefaultseppunct}\relax
\EndOfBibitem
\bibitem{Wang:2020prk}
G.-J. Wang {\em et~al.}, \ifthenelse{\boolean{articletitles}}{\emph{{Mass
  spectrum and strong decays of tetraquark $\cquarkbar\squarkbar \quark\quark$
  states}}, }{}\href{https://doi.org/10.1140/epjc/s10052-021-08978-0}{Eur.\
  Phys.\ J.\  \textbf{C81} (2021) 188},
  \href{http://arxiv.org/abs/2010.09395}{{\normalfont\ttfamily
  arXiv:2010.09395}}\relax
\mciteBstWouldAddEndPuncttrue
\mciteSetBstMidEndSepPunct{\mcitedefaultmidpunct}
{\mcitedefaultendpunct}{\mcitedefaultseppunct}\relax
\EndOfBibitem
\bibitem{Molina:2010tx}
R.~Molina, T.~Branz, and E.~Oset, \ifthenelse{\boolean{articletitles}}{\emph{{A
  new interpretation for the $D^*_{s2}(2573)$ and the prediction of novel
  exotic charmed mesons}},
  }{}\href{https://doi.org/10.1103/PhysRevD.82.014010}{Phys.\ Rev.\
  \textbf{D82} (2010) 014010},
  \href{http://arxiv.org/abs/1005.0335}{{\normalfont\ttfamily
  arXiv:1005.0335}}\relax
\mciteBstWouldAddEndPuncttrue
\mciteSetBstMidEndSepPunct{\mcitedefaultmidpunct}
{\mcitedefaultendpunct}{\mcitedefaultseppunct}\relax
\EndOfBibitem
\bibitem{Molina:2020hde}
R.~Molina and E.~Oset, \ifthenelse{\boolean{articletitles}}{\emph{{Molecular
  picture for the $X_0(2866)$ as a $D^* \bar{K}^*$ $J^P=0^+$ state and related
  $1^+$, $2^+$ states}},
  }{}\href{https://doi.org/10.1016/j.physletb.2020.135870}{Phys.\ Lett.\
  \textbf{B811} (2020) 135870},
  \href{http://arxiv.org/abs/2008.11171}{{\normalfont\ttfamily
  arXiv:2008.11171}}, [Erratum: Phys.Lett.B 837, 137645 (2023)]\relax
\mciteBstWouldAddEndPuncttrue
\mciteSetBstMidEndSepPunct{\mcitedefaultmidpunct}
{\mcitedefaultendpunct}{\mcitedefaultseppunct}\relax
\EndOfBibitem
\bibitem{Hu:2020mxp}
M.-W. Hu, X.-Y. Lao, P.~Ling, and Q.~Wang,
  \ifthenelse{\boolean{articletitles}}{\emph{{$X_0(2900)$ and its heavy quark
  spin partners in molecular picture}},
  }{}\href{https://doi.org/10.1088/1674-1137/abcfaa}{Chin.\ Phys.\
  \textbf{C45} (2021) 021003},
  \href{http://arxiv.org/abs/2008.06894}{{\normalfont\ttfamily
  arXiv:2008.06894}}\relax
\mciteBstWouldAddEndPuncttrue
\mciteSetBstMidEndSepPunct{\mcitedefaultmidpunct}
{\mcitedefaultendpunct}{\mcitedefaultseppunct}\relax
\EndOfBibitem
\bibitem{Liu:2020nil}
M.-Z. Liu, J.-J. Xie, and L.-S. Geng,
  \ifthenelse{\boolean{articletitles}}{\emph{{$X_0(2866)$ as a $\Dstar\Kstarb$
  molecular state}},
  }{}\href{https://doi.org/10.1103/PhysRevD.102.091502}{Phys.\ Rev.\
  \textbf{D102} (2020) 091502},
  \href{http://arxiv.org/abs/2008.07389}{{\normalfont\ttfamily
  arXiv:2008.07389}}\relax
\mciteBstWouldAddEndPuncttrue
\mciteSetBstMidEndSepPunct{\mcitedefaultmidpunct}
{\mcitedefaultendpunct}{\mcitedefaultseppunct}\relax
\EndOfBibitem
\bibitem{Kong:2021ohg}
S.-Y. Kong, J.-T. Zhu, D.~Song, and J.~He,
  \ifthenelse{\boolean{articletitles}}{\emph{{Heavy-strange meson molecules and
  possible candidates $D_{s0}^{*}(2317)$, $D_{s1}(2460)$, and $X_0(2900)$}},
  }{}\href{https://doi.org/10.1103/PhysRevD.104.094012}{Phys.\ Rev.\
  \textbf{D104} (2021) 094012},
  \href{http://arxiv.org/abs/2106.07272}{{\normalfont\ttfamily
  arXiv:2106.07272}}\relax
\mciteBstWouldAddEndPuncttrue
\mciteSetBstMidEndSepPunct{\mcitedefaultmidpunct}
{\mcitedefaultendpunct}{\mcitedefaultseppunct}\relax
\EndOfBibitem
\bibitem{Wang:2021lwy}
B.~Wang and S.-L. Zhu, \ifthenelse{\boolean{articletitles}}{\emph{{How to
  understand the $X(2900)$?}},
  }{}\href{https://doi.org/10.1140/epjc/s10052-022-10396-9}{Eur.\ Phys.\ J.\
  \textbf{C82} (2022) 419},
  \href{http://arxiv.org/abs/2107.09275}{{\normalfont\ttfamily
  arXiv:2107.09275}}\relax
\mciteBstWouldAddEndPuncttrue
\mciteSetBstMidEndSepPunct{\mcitedefaultmidpunct}
{\mcitedefaultendpunct}{\mcitedefaultseppunct}\relax
\EndOfBibitem
\bibitem{Xiao:2020ltm}
C.-J. Xiao, D.-Y. Chen, Y.-B. Dong, and G.-W. Meng,
  \ifthenelse{\boolean{articletitles}}{\emph{{Study of the decays of $S$-wave
  $\Dstarb \Kstar$ hadronic molecules: the scalar $X_0(2900)$ and its spin
  partners $\tilde{X}_{J(J=1,2)}$}},
  }{}\href{https://doi.org/10.1103/PhysRevD.103.034004}{Phys.\ Rev.\
  \textbf{D103} (2021) 034004},
  \href{http://arxiv.org/abs/2009.14538}{{\normalfont\ttfamily
  arXiv:2009.14538}}\relax
\mciteBstWouldAddEndPuncttrue
\mciteSetBstMidEndSepPunct{\mcitedefaultmidpunct}
{\mcitedefaultendpunct}{\mcitedefaultseppunct}\relax
\EndOfBibitem
\bibitem{Chen:2020aos}
H.-X. Chen, W.~Chen, R.-R. Dong, and N.~Su,
  \ifthenelse{\boolean{articletitles}}{\emph{{$X_0(2900)$ and $X_1(2900)$:
  hadronic molecules or compact tetraquarks}},
  }{}\href{https://doi.org/10.1088/0256-307X/37/10/101201}{Chin.\ Phys.\ Lett.\
   \textbf{37} (2020) 101201},
  \href{http://arxiv.org/abs/2008.07516}{{\normalfont\ttfamily
  arXiv:2008.07516}}\relax
\mciteBstWouldAddEndPuncttrue
\mciteSetBstMidEndSepPunct{\mcitedefaultmidpunct}
{\mcitedefaultendpunct}{\mcitedefaultseppunct}\relax
\EndOfBibitem
\bibitem{Chen:2021xlu}
H.-X. Chen, \ifthenelse{\boolean{articletitles}}{\emph{{Covalent hadronic
  molecules induced by shared light quarks}},
  }{}\href{https://doi.org/10.1088/1572-9494/ac8d8b}{Commun.\ Theor.\ Phys.\
  \textbf{74} (2022) 125201},
  \href{http://arxiv.org/abs/2105.09193}{{\normalfont\ttfamily
  arXiv:2105.09193}}\relax
\mciteBstWouldAddEndPuncttrue
\mciteSetBstMidEndSepPunct{\mcitedefaultmidpunct}
{\mcitedefaultendpunct}{\mcitedefaultseppunct}\relax
\EndOfBibitem
\bibitem{Huang:2020ptc}
Y.~Huang, J.-X. Lu, J.-J. Xie, and L.-S. Geng,
  \ifthenelse{\boolean{articletitles}}{\emph{{Strong decays of $\Dstarb \Kstar$
  molecules and the newly observed $X_{0,1}$ states}},
  }{}\href{https://doi.org/10.1140/epjc/s10052-020-08516-4}{Eur.\ Phys.\ J.\
  \textbf{C80} (2020) 973},
  \href{http://arxiv.org/abs/2008.07959}{{\normalfont\ttfamily
  arXiv:2008.07959}}\relax
\mciteBstWouldAddEndPuncttrue
\mciteSetBstMidEndSepPunct{\mcitedefaultmidpunct}
{\mcitedefaultendpunct}{\mcitedefaultseppunct}\relax
\EndOfBibitem
\bibitem{Burns:2020xne}
T.~J. Burns and E.~S. Swanson,
  \ifthenelse{\boolean{articletitles}}{\emph{{Discriminating among
  interpretations for the $X(2900)$ states}},
  }{}\href{https://doi.org/10.1103/PhysRevD.103.014004}{Phys.\ Rev.\
  \textbf{D103} (2021) 014004},
  \href{http://arxiv.org/abs/2009.05352}{{\normalfont\ttfamily
  arXiv:2009.05352}}\relax
\mciteBstWouldAddEndPuncttrue
\mciteSetBstMidEndSepPunct{\mcitedefaultmidpunct}
{\mcitedefaultendpunct}{\mcitedefaultseppunct}\relax
\EndOfBibitem
\bibitem{Qin:2022nof}
Q.~Qin, J.-L. Qiu, and F.-S. Yu,
  \ifthenelse{\boolean{articletitles}}{\emph{{Diagrammatic analysis of hidden-
  and open-charm tetraquark production in $B$ decays}},
  }{}\href{https://doi.org/10.1140/epjc/s10052-023-11375-4}{Eur.\ Phys.\ J.\ C
  \textbf{83} (2023) 227},
  \href{http://arxiv.org/abs/2212.03590}{{\normalfont\ttfamily
  arXiv:2212.03590}}\relax
\mciteBstWouldAddEndPuncttrue
\mciteSetBstMidEndSepPunct{\mcitedefaultmidpunct}
{\mcitedefaultendpunct}{\mcitedefaultseppunct}\relax
\EndOfBibitem
\bibitem{Chen:2022asf}
H.-X. Chen {\em et~al.}, \ifthenelse{\boolean{articletitles}}{\emph{{An updated
  review of the new hadron states}},
  }{}\href{https://doi.org/10.1088/1361-6633/aca3b6}{Rept.\ Prog.\ Phys.\
  \textbf{86} (2023) 026201},
  \href{http://arxiv.org/abs/2204.02649}{{\normalfont\ttfamily
  arXiv:2204.02649}}\relax
\mciteBstWouldAddEndPuncttrue
\mciteSetBstMidEndSepPunct{\mcitedefaultmidpunct}
{\mcitedefaultendpunct}{\mcitedefaultseppunct}\relax
\EndOfBibitem
\bibitem{LHCb-DP-2008-001}
LHCb collaboration, A.~A. Alves~Jr.\ {\em et~al.},
  \ifthenelse{\boolean{articletitles}}{\emph{{The \lhcb detector at the LHC}},
  }{}\href{https://doi.org/10.1088/1748-0221/3/08/S08005}{JINST \textbf{3}
  (2008) S08005}\relax
\mciteBstWouldAddEndPuncttrue
\mciteSetBstMidEndSepPunct{\mcitedefaultmidpunct}
{\mcitedefaultendpunct}{\mcitedefaultseppunct}\relax
\EndOfBibitem
\bibitem{LHCb-DP-2014-002}
LHCb collaboration, R.~Aaij {\em et~al.},
  \ifthenelse{\boolean{articletitles}}{\emph{{LHCb detector performance}},
  }{}\href{https://doi.org/10.1142/S0217751X15300227}{Int.\ J.\ Mod.\ Phys.\
  \textbf{A30} (2015) 1530022},
  \href{http://arxiv.org/abs/1412.6352}{{\normalfont\ttfamily
  arXiv:1412.6352}}\relax
\mciteBstWouldAddEndPuncttrue
\mciteSetBstMidEndSepPunct{\mcitedefaultmidpunct}
{\mcitedefaultendpunct}{\mcitedefaultseppunct}\relax
\EndOfBibitem
\bibitem{PDG2024}
Particle Data Group, S.~Navas {\em et~al.},
  \ifthenelse{\boolean{articletitles}}{\emph{{\href{http://pdg.lbl.gov/}{Review
  of particle physics}}},
  }{}\href{https://doi.org/10.1103/PhysRevD.110.030001}{Phys.\ Rev.\
  \textbf{D110} (2024) 030001}\relax
\mciteBstWouldAddEndPuncttrue
\mciteSetBstMidEndSepPunct{\mcitedefaultmidpunct}
{\mcitedefaultendpunct}{\mcitedefaultseppunct}\relax
\EndOfBibitem
\bibitem{Hulsbergen:2005pu}
W.~D. Hulsbergen, \ifthenelse{\boolean{articletitles}}{\emph{{Decay chain
  fitting with a Kalman filter}},
  }{}\href{https://doi.org/10.1016/j.nima.2005.06.078}{Nucl.\ Instrum.\ Meth.\
  \textbf{A552} (2005) 566},
  \href{http://arxiv.org/abs/physics/0503191}{{\normalfont\ttfamily
  arXiv:physics/0503191}}\relax
\mciteBstWouldAddEndPuncttrue
\mciteSetBstMidEndSepPunct{\mcitedefaultmidpunct}
{\mcitedefaultendpunct}{\mcitedefaultseppunct}\relax
\EndOfBibitem
\bibitem{Breiman}
L.~Breiman, J.~H. Friedman, R.~A. Olshen, and C.~J. Stone, {\em Classification
  and regression trees}, Wadsworth international group, Belmont, California,
  USA, 1984\relax
\mciteBstWouldAddEndPuncttrue
\mciteSetBstMidEndSepPunct{\mcitedefaultmidpunct}
{\mcitedefaultendpunct}{\mcitedefaultseppunct}\relax
\EndOfBibitem
\bibitem{AdaBoost}
Y.~Freund and R.~E. Schapire, \ifthenelse{\boolean{articletitles}}{\emph{A
  decision-theoretic generalization of on-line learning and an application to
  boosting}, }{}\href{https://doi.org/10.1006/jcss.1997.1504}{J.\ Comput.\
  Syst.\ Sci.\  \textbf{55} (1997) 119}\relax
\mciteBstWouldAddEndPuncttrue
\mciteSetBstMidEndSepPunct{\mcitedefaultmidpunct}
{\mcitedefaultendpunct}{\mcitedefaultseppunct}\relax
\EndOfBibitem
\bibitem{Hocker:2007ht}
H.~Voss, A.~Hoecker, J.~Stelzer, and F.~Tegenfeldt,
  \ifthenelse{\boolean{articletitles}}{\emph{{TMVA - Toolkit for Multivariate
  Data Analysis with ROOT}}, }{}\href{https://doi.org/10.22323/1.050.0040}{PoS
  \textbf{ACAT} (2007) 040}\relax
\mciteBstWouldAddEndPuncttrue
\mciteSetBstMidEndSepPunct{\mcitedefaultmidpunct}
{\mcitedefaultendpunct}{\mcitedefaultseppunct}\relax
\EndOfBibitem
\bibitem{TMVA4}
A.~Hoecker {\em et~al.}, \ifthenelse{\boolean{articletitles}}{\emph{{TMVA 4 ---
  Toolkit for Multivariate Data Analysis with ROOT. Users Guide.}},
  }{}\href{http://arxiv.org/abs/physics/0703039}{{\normalfont\ttfamily
  arXiv:physics/0703039}}\relax
\mciteBstWouldAddEndPuncttrue
\mciteSetBstMidEndSepPunct{\mcitedefaultmidpunct}
{\mcitedefaultendpunct}{\mcitedefaultseppunct}\relax
\EndOfBibitem
\bibitem{Sjostrand:2007gs}
T.~Sj\"{o}strand, S.~Mrenna, and P.~Skands,
  \ifthenelse{\boolean{articletitles}}{\emph{{A brief introduction to PYTHIA
  8.1}}, }{}\href{https://doi.org/10.1016/j.cpc.2008.01.036}{Comput.\ Phys.\
  Commun.\  \textbf{178} (2008) 852},
  \href{http://arxiv.org/abs/0710.3820}{{\normalfont\ttfamily
  arXiv:0710.3820}}\relax
\mciteBstWouldAddEndPuncttrue
\mciteSetBstMidEndSepPunct{\mcitedefaultmidpunct}
{\mcitedefaultendpunct}{\mcitedefaultseppunct}\relax
\EndOfBibitem
\bibitem{Lange:2001uf}
D.~J. Lange, \ifthenelse{\boolean{articletitles}}{\emph{{The EvtGen particle
  decay simulation package}},
  }{}\href{https://doi.org/10.1016/S0168-9002(01)00089-4}{Nucl.\ Instrum.\
  Meth.\  \textbf{A462} (2001) 152}\relax
\mciteBstWouldAddEndPuncttrue
\mciteSetBstMidEndSepPunct{\mcitedefaultmidpunct}
{\mcitedefaultendpunct}{\mcitedefaultseppunct}\relax
\EndOfBibitem
\bibitem{Allison:2006ve}
Geant4 collaboration, J.~Allison {\em et~al.},
  \ifthenelse{\boolean{articletitles}}{\emph{{Geant4 developments and
  applications}}, }{}\href{https://doi.org/10.1109/TNS.2006.869826}{IEEE
  Trans.\ Nucl.\ Sci.\  \textbf{53} (2006) 270}\relax
\mciteBstWouldAddEndPuncttrue
\mciteSetBstMidEndSepPunct{\mcitedefaultmidpunct}
{\mcitedefaultendpunct}{\mcitedefaultseppunct}\relax
\EndOfBibitem
\bibitem{Muller:2018vny}
D.~M\"uller, M.~Clemencic, G.~Corti, and M.~Gersabeck,
  \ifthenelse{\boolean{articletitles}}{\emph{{ReDecay: A novel approach to
  speed up the simulation at LHCb}},
  }{}\href{https://doi.org/10.1140/epjc/s10052-018-6469-6}{Eur.\ Phys.\ J.\
  \textbf{C78} (2018) 1009},
  \href{http://arxiv.org/abs/1810.10362}{{\normalfont\ttfamily
  arXiv:1810.10362}}\relax
\mciteBstWouldAddEndPuncttrue
\mciteSetBstMidEndSepPunct{\mcitedefaultmidpunct}
{\mcitedefaultendpunct}{\mcitedefaultseppunct}\relax
\EndOfBibitem
\bibitem{JACOB1959404}
M.~Jacob and G.~C. Wick, \ifthenelse{\boolean{articletitles}}{\emph{{On the
  general theory of collisions for particles with spin}},
  }{}\href{https://doi.org/10.1006/aphy.2000.6022}{Annals Phys.\  \textbf{7}
  (1959) 404}\relax
\mciteBstWouldAddEndPuncttrue
\mciteSetBstMidEndSepPunct{\mcitedefaultmidpunct}
{\mcitedefaultendpunct}{\mcitedefaultseppunct}\relax
\EndOfBibitem
\bibitem{Back:2017zqt}
J.~Back {\em et~al.}, \ifthenelse{\boolean{articletitles}}{\emph{{{\sc
  Laura}$^{++}$: A Dalitz plot fitter}},
  }{}\href{https://doi.org/10.1016/j.cpc.2018.04.017}{Comput.\ Phys.\ Commun.\
  \textbf{231} (2018) 198},
  \href{http://arxiv.org/abs/1711.09854}{{\normalfont\ttfamily
  arXiv:1711.09854}}\relax
\mciteBstWouldAddEndPuncttrue
\mciteSetBstMidEndSepPunct{\mcitedefaultmidpunct}
{\mcitedefaultendpunct}{\mcitedefaultseppunct}\relax
\EndOfBibitem
\bibitem{Chung:1995dx}
S.~U. Chung {\em et~al.}, \ifthenelse{\boolean{articletitles}}{\emph{{Partial
  wave analysis in K-matrix formalism}},
  }{}\href{https://doi.org/10.1002/andp.19955070504}{Annalen Phys.\  \textbf{4}
  (1995) 404}\relax
\mciteBstWouldAddEndPuncttrue
\mciteSetBstMidEndSepPunct{\mcitedefaultmidpunct}
{\mcitedefaultendpunct}{\mcitedefaultseppunct}\relax
\EndOfBibitem
\end{mcitethebibliography}

\newpage
\centerline
{\large\bf LHCb collaboration}
\begin
{flushleft}
\small
R.~Aaij$^{38}$\lhcborcid{0000-0003-0533-1952},
A.S.W.~Abdelmotteleb$^{57}$\lhcborcid{0000-0001-7905-0542},
C.~Abellan~Beteta$^{51}$,
F.~Abudin{\'e}n$^{57}$\lhcborcid{0000-0002-6737-3528},
T.~Ackernley$^{61}$\lhcborcid{0000-0002-5951-3498},
A. A. ~Adefisoye$^{69}$\lhcborcid{0000-0003-2448-1550},
B.~Adeva$^{47}$\lhcborcid{0000-0001-9756-3712},
M.~Adinolfi$^{55}$\lhcborcid{0000-0002-1326-1264},
P.~Adlarson$^{82}$\lhcborcid{0000-0001-6280-3851},
C.~Agapopoulou$^{14}$\lhcborcid{0000-0002-2368-0147}
% ...

{\footnotesize \it

% ...
$^{14}$Universit{\'e} Paris-Saclay, CNRS/IN2P3, IJCLab, Orsay, France\\
% ...
$^{38}$Nikhef National Institute for Subatomic Physics, Amsterdam, Netherlands\\
% ...
$^{47}$Instituto Galego de F{\'\i}sica de Altas Enerx{\'\i}as (IGFAE), Universidade de Santiago de Compostela, Santiago de Compostela, Spain\\
% ...
$^{51}$Physik-Institut, Universit{\"a}t Z{\"u}rich, Z{\"u}rich, Switzerland\\
% ...
$^{55}$H.H. Wills Physics Laboratory, University of Bristol, Bristol, United Kingdom\\
% ...
$^{57}$Department of Physics, University of Warwick, Coventry, United Kingdom\\
% ...
$^{60}$School of Physics and Astronomy, University of Glasgow, Glasgow, United Kingdom\\
$^{61}$Oliver Lodge Laboratory, University of Liverpool, Liverpool, United Kingdom\\
% ...
$^{69}$Syracuse University, Syracuse, NY, United States\\
% ...
$^{82}$Department of Physics and Astronomy, Uppsala University, Uppsala, Sweden, associated to $^{60}$\\
% ...
}
\end{flushleft}

\end{document}